\documentclass[aps,prl,twocolumn,superscriptaddress]{revtex4-1}

\usepackage{graphicx}% Include figure files
\usepackage{dcolumn}% Align table columns on decimal point
\usepackage{bm}% bold math
\usepackage{amsmath}
\usepackage{multirow}
\usepackage{color}

{\catcode `\@=11 \global\let\AddToReset=\@addtoreset}
\AddToReset{equation}{section}

\newcounter{mnotecount}[section]
\renewcommand{\themnotecount}{\thesection.\arabic{mnotecount}}
\newcommand{\mnotex}[1]%{}
{\protect{\stepcounter{mnotecount}}$^{\mbox{\footnotesize
$%\!\!\!\!\!\!\,
\bullet$\themnotecount}}$ \marginpar{%\color{red}%
\raggedright\tiny\em
$\!\!\!\!\!\!\,\bullet$\themnotecount: #1} }

\usepackage[caption=false]{subfig}

\usepackage{chngcntr}
\counterwithout{equation}{section}

\begin{document}

\title{Direct measurement of short-range forces with a levitated nanoparticle}

\author{George Winstone} 
\affiliation{Department of Physics and Astronomy, University of Southampton, Southampton SO17 1BJ, UK}
\affiliation{School for Materials Science, Japan Advanced Institute of Science and Technology, Nomi, Ishikawa 923-1211, Japan}

\author{Markus Rademacher}
\affiliation{Department of Physics and Astronomy, University of Southampton, Southampton SO17 1BJ, UK}
 
\author{Robert Bennett}
\affiliation{Institute of Physics, Albert-Ludwigs-University Freiburg, D-79104 Freiburg, Germany}

\author{Stefan Buhmann}
\affiliation{Institute of Physics, Albert-Ludwigs-University Freiburg, D-79104 Freiburg, Germany}
\affiliation{Freiburg Institute for Advanced Studies (FRIAS), D-79104 Freiburg, Germany}

\author{Hendrik Ulbricht}
\email[]{h.ulbricht@soton.ac.uk}
\affiliation{Department of Physics and Astronomy, University of Southampton, Southampton SO17 1BJ, UK}

\date{\today}
            
%\begin{abstract}
%We report on optical levitation experiments to probe the interaction of a nanoparticle with a surface in vacuum. The observed interaction-induced effect is a controllable anharmonicity of the particle trapping potential. We reconstruct the Coulomb image charge interaction potential to be in perfect agreement with the experimental data. We find while the particle in our experiment is electrically charged, the force sensitivity we reach in the experiment is able to resolve the Casimir-Polder interaction. The results may open the possibility for a new surface sensitive scanning probe technique based on the extremely high mechanical sensitivity of levitated nanoparticles. 
%\end{abstract}

%\pacs{Valid PACS appear here}

%%Intro

\begin{abstract}
Short-range forces have important real-world relevance across a range of settings in the nano world, from colloids \cite{Lee2001} and possibly for protein folding \cite{dill1990dominant, nicholls1991protein, yang2013much} to nano-mechanical devices \cite{RevModPhys.82.1887, Geraci2010Short-rangeMicrospheres}, but also for detection of weak long-range forces, such as gravity, at short distances \cite{Kapner2007, Geraci2010Short-rangeMicrospheres, Moore2014, schmole2016micromechanical} and of candidates to solve the problem of dark energy \cite{hamilton2015atom}. Short-range forces, such as Casimir-Polder or van der Waals are in general difficult to calculate as a consequence of their non-additive nature, and challenging to measure due to their small magnitude --- especially for charged particles where dispersion forces are normally many orders of magnitude smaller than electrostatic image forces. Therefore short-range forces have represented a continuing theoretical and experimental challenge over the last half-century \cite{Shih1975, Sukenik1993, Obrecht2007, bender2014,Lamoreaux1997,Mohideen1998}.
Here we report on experiments with a single glass nanoparticle levitated in close proximity to a neutral silicon surface in vacuum, which allow for direct measurement of short-range forces in a new distance and sensitivity regime - outperforming existing force microscopies \cite{PhysRevLett.56.930}. 
\end{abstract}

\maketitle

\begin{figure*}
\includegraphics[width=1.8\columnwidth]{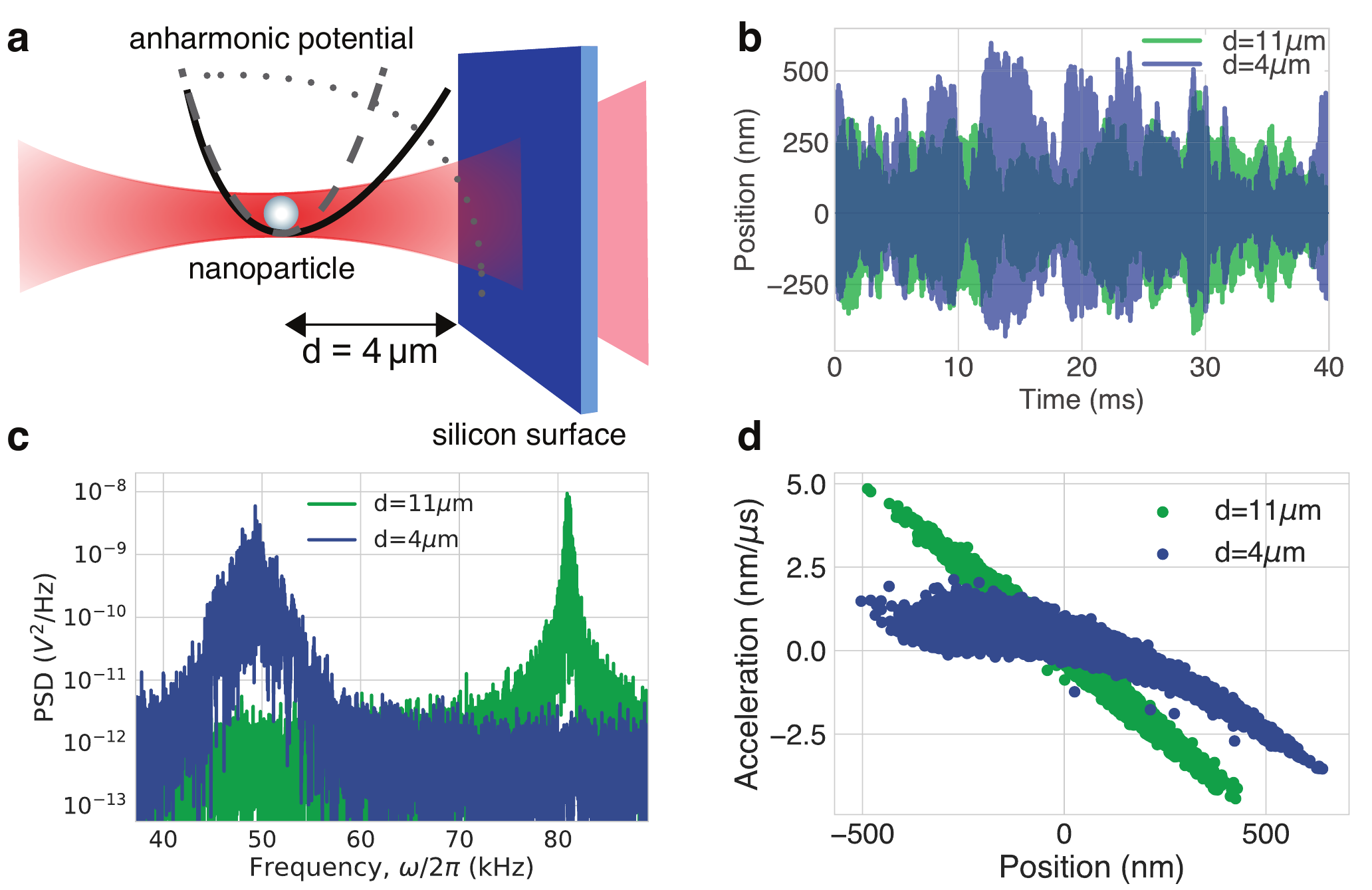}
\caption{\textbf{Particle-surface experiment with a levitated nanosphere.} (a) Schematic of the experiment. The particle is optically trapped close to the surface at various distances $d$. (b) Time-trace of the particle position trapped close and far away from the surface. The amplitude of the oscillation grows when the particle is closer to the surface. (c) Power spectral density (PSD) of $z$-motion of the trapped particle at two different distances, and (d) the related spring functions. Spring function shows non-linear shape, if the particle is close to the surface.}
\label{fig:surface_conceptual}
\end{figure*}

A charged particle near a surface will experience an attractive short-range force due to the interaction with its image charge. A competing class of effects that persist even for neutral objects arise from correlations between the fluctuations of atomic dipoles that make up two spatially-separated bodies --- these are dispersion forces, known as the Casimir force \cite{CasimirOriginal1948} for macroscopic objects, and if one of them is microscopic (atom, molecule, nanosphere, etc) then the resulting effects are variously termed Casimir-Polder \cite{Casimir1948} or van der Waals forces. A model system in which to study short-range forces is two closely-spaced objects separated by vacuum, as depicted in figure \ref{fig:surface_conceptual}, where the particle-surface interaction is probed for varying distances. Over the years a variety of experiments have been performed with different physical systems (e.g. torsion pendulums \cite{Lamoreaux1997}, cantilevers, and tip probes \cite{Mohideen1998}) to investigate surface forces. Several experiments that utilize controllable cold atoms have been performed to measure Casimir-Polder and van der Waals forces close to an uncharged surface \cite{Shih1975, Sukenik1993, Obrecht2007, bender2014} 

\begin{figure*}
\includegraphics[width=1.8\columnwidth]{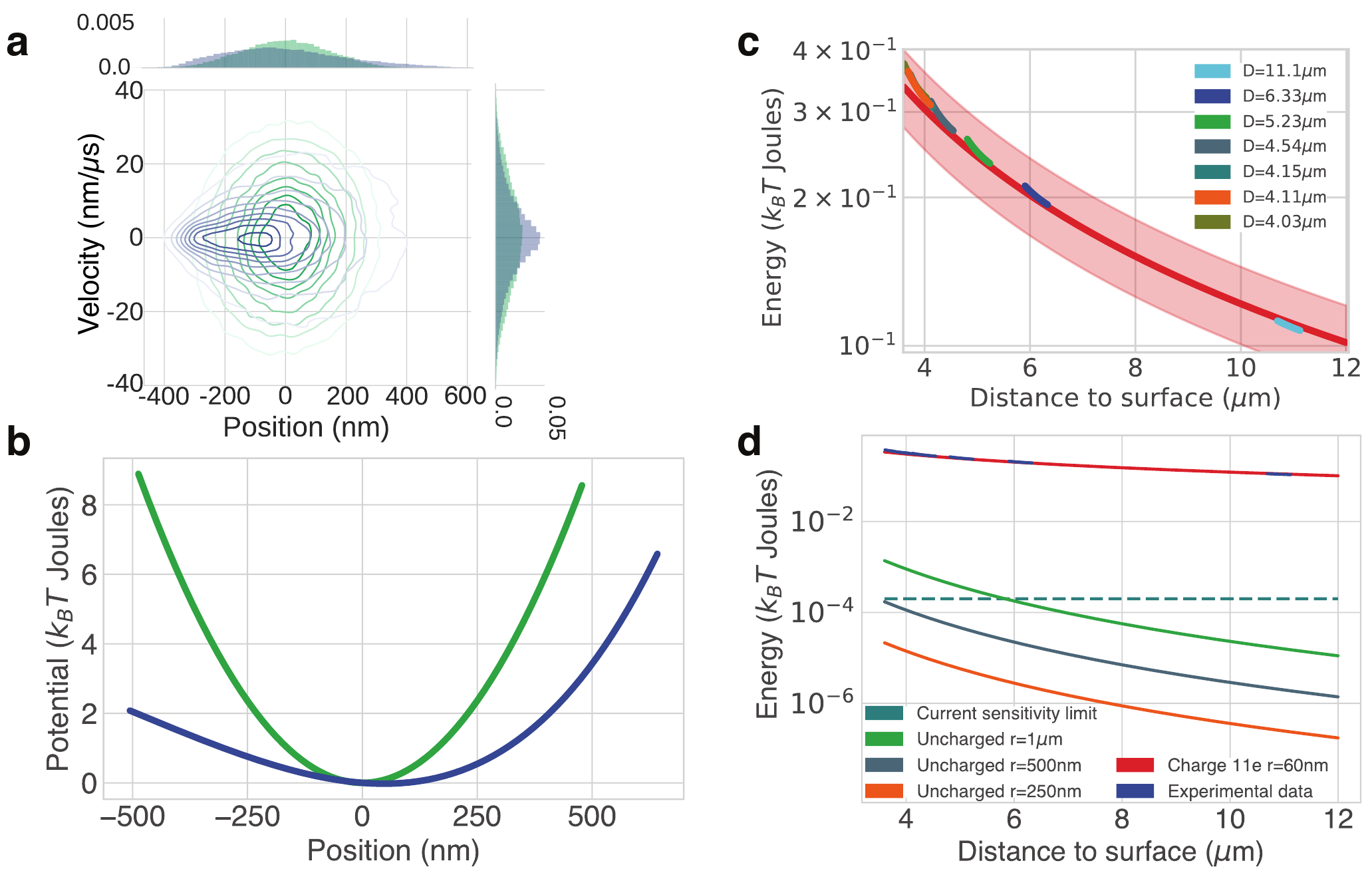}
\caption{\textbf{The nanoparticle surface probe.} (a) shows the dynamics of the particle in phase-space representation. Contours are experimental data for positions and velocity of the particle close (blue) and far away (green) from the surface. Clearly visible is the non-spherical shape for the case of the particle close to the surface which is due to the interaction with the surface. (b) The experimental reconstructed potential as experienced by the particle at the different distances. The potential becomes anharmonic if the particle is closer to the surface. (c) Compares the experimental data taken at seven different distances and the theory according to equation(\ref{mirrorchargepotential}), red line.  The pink region indicated the same mirror charge interaction by with $\pm$1 elementary charge $e$. Closer to the surface the best fit with -11$e$ deviates from the experimental data. (d) Comparison of experimental interaction data with different types of potentials, such as Casimir-Polder for particles of different size as well as equation(\ref{mirrorchargepotential}). The horizontal dashed line is the sensitivity limit of the present experiment.}
\label{fig:experimentalsetupsurface}
\end{figure*}

The experimental configuration is depicted schematically in figure \ref{fig:surface_conceptual}a). A 60 nm radius silica nanoparticle is trapped in a tiny light spot focused by a parabolic mirror close to a silicon surface (a 200 $\mu$m thick highly n-doped Si(100) wafer with a 300 nm SiO$_2$ layer on top), the particle-surface distance $d$ is varied between 4 $\mu$m and 11 $\mu$m at vacuum of 10$^{-2}$mbar. At such pressure the motion of the particle is still affected by stochastic background gas collisions and we give the equation describing the dynamics of the particle under that circumstance in the methods section, however here we concentrate on a different aspect of the motion. The particle position is not stabilised by any feedback and is allowed to move freely in the trap at different $d$. The particle is electrically charged and we evaluate that it carries a charge of eleven elementary charges $e$. Based on our optical detection of the particles position we can measure time traces of the particle position with high interferometric resolution of 1 pm. Typical time-domain data of the trapped particle are shown in figure \ref{fig:surface_conceptual}b). These data contain information about the motion of the particle in all three spatial directions, however we can separate those in the frequency domain and concentrate here only on the motion in $z$-direction, which is normal to the plane of the trapping mirror.  The power spectral density (PSD) of the $z$-motion is shown in figure \ref{fig:surface_conceptual}c). From such data we can extract the actual shape of the trapping potential, which is harmonic for the optical trap of the particle. If the particle is close to the surface that potential is affected, in particular the potential becomes anharmonic, which we can directly extract form the data as shown in figure \ref{fig:experimentalsetupsurface}b). We then reconstruct the surface potential from that experimental data, which is the basic technique of this paper. We can also extract the so-called spring function of the motion of the particle, which shows a clear distinction for close and far away from the surface, see figure \ref{fig:surface_conceptual}d).

%%%%%Theory

At large distances from the surface, where the surface is not affecting the oscillation of the particle, the levitated nanoparticle is trapped optically in the focus of a Gaussian laser beam in the Rayleigh limit, within an optical harmonic potential $U_0(x)= (k/2) x^2$, where $x$ is the spatial displacement of the nanosphere while oscillating, and $k$ is the spring constant, which for the optical trap is originated by the optical gradient force and therefore $k=2\alpha P /(c\pi\epsilon_0 w^6/\lambda^2)$, with $\alpha$ being the polarisability of the nanoparticle, $P$ the incident laser power,  $c$ the speed of light, $\epsilon_0$ the permittivity of free space, $w$ is the laser waist at focus, and $\lambda$ the wavelength of the laser. Here we only consider the one-dimensional $z$-motion of the particle normal to the surface. More details about the optical trap can be found elsewhere \cite{Rashid2016}.

The potential of a charged particle interacting with its image charge $U_\text{ic}(d)$ in a dielectric substrate with a layer of thickness $L$ deposited on top is;
\begin{equation}
U_\text{ic}(d)=-\frac{Q^2}{4\pi\epsilon_0}\frac{1}{2} \int_0^\infty dz \frac{R_{01} +R_{12}e^{-2 z L}}{1+R_{01}R_{12}e^{-2 z L}}e^{-2 z d} ,
\label{mirrorchargepotential}
\end{equation}
with $R_{01} = (\epsilon_1-1)/(\epsilon_1+1), R_{12} = (\epsilon_2-\epsilon_1)/(\epsilon_2+\epsilon_1)$, where $\epsilon_1$ and $\epsilon_2$ are the permittivities of the layer and the substrate respectively. $Q$ is the charge of the particle and $d$ is the distance between the nanoparticle and the vacuum-layer interface, and $z$ is the integration variable. Close to the surface the particle will explore a total potential  $U_{t}(d)$ defined by the superposition of the surface interaction potential and the optical potential, $U_\text{t}(d) = U_\text{ic}(d) + \frac{1}{2}k (x-d)^2$.

The total potential is now an anharmonic potential, where the non-linearity is added through the image charge surface interaction in addition to the harmonic potential of the optical trap.

Aside from the Coulomb potential, the nanosphere experiences a dispersion force arising from correlations between the fluctuations of its own atomic dipoles and those in the surface. Here, the nanosphere is far enough away from the surface that it can be considered as a point dipole, with polarisability $\alpha(\omega)$ obtained from the well-known Clausius-Mossotti relation for a sphere of radius $R$ and permittivity $\epsilon(\omega)$; $\alpha(\omega) = 4 \pi \epsilon_0 R^3 (\epsilon(\omega)-1)/(\epsilon(\omega)+2)$.

In the supplementary material we present numerical results for the Casimir-Polder potential for a wide range of distances, however in order to gain a simple and useable formula we note that the nanoparticle-surface distance (4-11 $\mu$m) is large compared to the wavelength of any of the dominant transitions in the optical response of either of the materials (70 nm for SiO$_2$ and 265 nm for Si).  Thus we are in the retarded regime, where the Casimir-Polder potential has the form \cite{Palik1985};
\begin{equation}
U_\text{CP}(d) = -\frac{C_4}{d^4}
\end{equation}   
where $C_4$ is a distance-independent constant defined as \cite{Spruch1993};
\begin{equation}\label{C4Def}
C_4\! = \!\frac{3\hbar c  \alpha(0)}{64\pi^2 \epsilon_0} \!\!\int_1^\infty\! \!\!dv \left(\frac{2}{v^2}\!-\!\frac{1}{v^4}\right) \frac{\varepsilon_1(0) v-\sqrt{\varepsilon_1(0) -1+v^2}}{\varepsilon_1(0) v+\sqrt{\varepsilon_1(0) -1+v^2}}
\end{equation}
Using the measured optical data for silicon and silicon dioxide presented in the Supplementary Material as tabulated in \cite{Palik1985}, we find a value of;
$C_4= (7.60\times10^{-28}\text{Jm})\cdot R^3$. In the following we compare the data to the trapping and Coulomb potentials $U_\text{ic}$ and $U_0$.

%%%%Results

The mass of the nanoparticle is extracted by comparing the potential for the steady state Langevin equation result for our system with the potential obtained by integrating the particle's position-acceleration relation (see figure \ref{fig:surface_conceptual}d)). The radius is extracted from this based on the assumption that the particle is of spherical shape. Then the radius of the nanoparticle in the experiments was extracted from experimental data to be $r=$60 nm ($\pm$5 nm). More details about the procedure for particle size estimation is described in the supplement. 

Figure \ref{fig:experimentalsetupsurface}a) shows the phase space of the nanosphere's centre of mass motion far away from the surface (green), and at the closest available position, $4 \mu$m (blue), before the surface forces overpower the optical forces and the particle gets lost from the trap. The position distribution of the particle is drawn towards the surface and the motion becomes significantly anharmonic. We reconstruct the potential $U(r)$ at position $r$ from time-domain position measurements by calculating the acceleration or spring function of the particle's motion, which is proportional to the force acting on the particle in the potential.  For a linear simple harmonic oscillator system the spring constant defines the relation between the acceleration and displacement of the test mass: $F = ma = kx$. In the case of a non-linear or anharmonic oscillator, the spring constant is often no longer a constant but a function of the displacement $k(x)$, leading to the functions for the linear and non-linear cases given in the supplement. Integrating the relation, $F(r)=-\nabla (U(r))$, with respect to $r$ then gives the reconstruction of the interaction potential. Similar methods have been used earlier \cite{Rondin2017}. In the case of a simple, steady state, differentiable potential, it is therefore possible to reconstruct the potential from the spring functions of the particle's motion at each distance. The spring functions for the particle with and without surface are shown in figure \ref{fig:springs} in the supplement. For experimental data, with and without the surface, we show the potentials reconstructed in figure \ref{fig:experimentalsetupsurface}b), and their corresponding spring functions in figure \ref{fig:surface_conceptual}d) and more details in the supplement. 

At smaller particle-surface distances $d$, the trapping potential experiences an increasingly strong perturbation from the surface interaction. Comparing the reconstructed potential with different interaction models shows the best agreement for the case of image charge interaction of charge $Q=$-11e ($\pm$1e) from fitting the equation \ref{mirrorchargepotential} to the experimental data, as shown in figure \ref{fig:experimentalsetupsurface}c). The observed particle net-charge is in agreement with typical values in recent experiments with trapped nanoparticles \cite{Frimmer2017, hempston2017force}. The observed deviation from the model at small distances could be attributable to electrostatic patch effects, see figure \ref{fig:experimentalsetupsurface}c). Charge and electric dipole patch effects have been shown to contribute in high-sensitivity surface force measurements \cite{sushkov2011observation, behunin2012electrostatic}.

To estimate the experimental sensitivity, we perturb a suspended nanoparticle with an electric field \cite{hempston2017force}. This allows us to resolve changes to the potential structure of $2\cdot10^{-4} k_B T$. Encouragingly, applying such a resolution to the distance ranges scanned in this experiment predicts we should be able to resolve Casimir Polder forces if the same surface-nanoparticle experiment was to be repeated with a larger particle.  We evaluate the experiment to be sensitive to surface forces of $10^{-19}$ N/$\sqrt{\textrm{Hz}}$, in contrast to $10^{-10}$ N/$\sqrt{\textrm{Hz}}$ in the original  atomic force microscope (AFM) paper \cite{PhysRevLett.56.930}. This level of force sensitivity allows for detection of genuine Casimir-Polder interactions for a particle of radius 1 $\mu$m, while with the best sensitivity demonstrated in this system to date \cite{hempston2017force}, the study of CP with a 300 nm radius particle appears to be within reach. Encouragingly, the smallest particle-surface {\it interaction energies} measured here are on the order of 100 $\mu$eV, which is the order of magnitude for dispersion forces - much smaller energies than those typical for covalent bonds and charge transfer interactions. The {\it spatial resolution} of position detection is given by the parabolic mirror trap detection technique and has been demonstrated to reach 200 fm/$\sqrt{\textrm{Hz}}$ \cite{vovrosh2017parametric}. This makes the spatial resolution of our surface probe technique much finer than the size of the trap, which is on the order of several hundred nm. 

\begin{figure}
\centering
\includegraphics[width=\columnwidth]{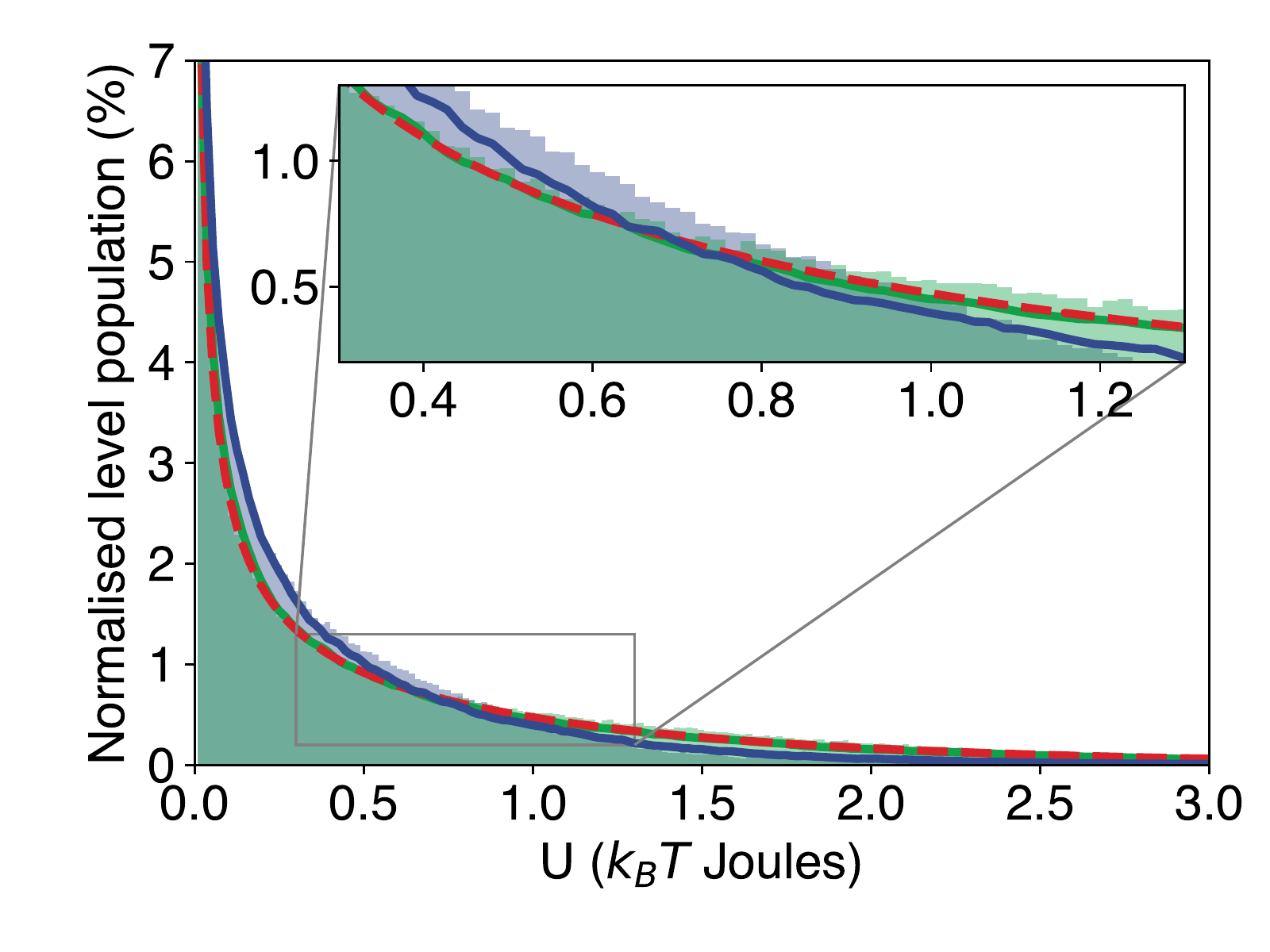}
\caption{{\bf Energy comparison: harmonic vs anharmonic trap.} Normalised histogram of the particle's potential energy for the cases of an harmonic potential (particle far away from surface, green) and an anharmonic Morse-like potential (particle close to the surface, blue). Shown in red are the results of numerical simulation of the particle's motion for harmonic (dashed) and anharmonic (dotted) potentials with $10 \, 000$ different realisations with randomised amplitudes. The solid blue line is the normalised result of equation~\eqref{dtAvg} in the supplement. The Morse-like potential shows a similar behaviour in energy level scaling as compared to the harmonic case until diverging strongly at a high value of n, which is the expected Morse-like behaviour. See supplement for computational details.
}
\label{fig:outlook_with_data}
\end{figure}

To probe further the Morse-type of the reconstructed anharmonic potential, we plot a histogram of the particle potential energy and compare again the two cases for the particle with and without surface. The plot is shown in figure \ref{fig:outlook_with_data} (and \ref{fig:morse_potential_level_behaviour} in the supplement) and the anharmonic potential energy distribution can be well-understood from numerically solving the equation of motion, see supplement for computational details.  

%%%%%Discussion

The use of a nanoparticle as opposed to an atom for force detection has the advantage that the dispersion coefficient $C_4$ is much larger, giving a stronger force. For example taking a selection of alkali metals  \cite{Derevianko1999} (commonly used in atom-surface experiments), one finds $C_4$ coefficients of around $10^{-56}$Jm$^{4}$, but for a nanosphere of radius $60$ nm the corresponding value from equation \eqref{C4Def} is $1.6\times 10^{-49}$Jm$^4$ --- seven orders of magnitude bigger. This difference can be qualitatively understood by noting that a hypothetical nanosphere with a radius of $1-2$\AA \,would result in approximately the same $C_4$ as an atom. The $60$ nm sphere in the experiment is around 300 times larger than this, so the cubic scaling of $C_4$ with the radius means that the nanosphere $C_4$ is a factor of around $300^6\sim 10^7$ larger, as reflected in the calculation above.\\

%%%%Conclusion

In summary, we have presented an experiment where the surface-induced force on a levitated nanoparticle can be directly observed. The particle-surface interaction induces an anharmonic trapping potential in deviation from the harmonic ($x^2$) behaviour and therefore generates a non-linearity in the motion of the particle. This non-linearity affects all trapped states, also such with small amplitudes which may be relevant for the use of such non-linearities to generate nonclassical motional states of nanoparticle optomechanics. We associate the observed anharmonic effect with a particle-surface interaction involving electric charge. In principle, the current parameters of force sensitivity and particle-surface distances allow for the detection of Casimir-Polder forces with a levitated nanoparticle if the experiment is repeated with a particle with zero net-charge and of reasonable size. The careful analysis of the particle position measured in the lateral directions along the surface may allow for topographic images in close analogy to scanning probe techniques. A further advantage of a levitated nanoparticle probe might be that the measurement can be performed with very high precision while both surface and nanoparticle are at room temperature. This could be interesting in the context of biological and physiological samples. As suggested recently, rotation of non-spherical nanoparticles close to surfaces might be another interesting system for investigation of surface forces \cite{Manjavacas2017}. This work may pave the way for cooling and trapping of nanoparticles in vacuum close to surfaces in self-induced back action type near-field traps \cite{juan2009self}, invoking both plasmonic or photonic crystal effects to trap the particle.

%{\bf Methods} 

%Methods and any associated references are available in the online version of the paper.

{\it Acknowledgements--} We would like to thank for discussions Muddassar Rashid, David Hempston, Marko Toro{\v{s}, Chris Timberlake, and Ashley Setter. H.U. and G.W. acknowledge funding by The Leverhulme Trust and the Foundational Questions Institute (FQXi). G.W. acknowledges the collaborative education and research co-supervision program between University of Southampton and JAIST. M.R. would like to acknowledge funding by the Erasmus+ program by \"{O}sterreichischer Austauschdienst and the Top-Stipendium exchange scholarship of the state of Lower Austria. R.B. and S.B. acknowledge funding by Deutsche Forschungsgemeinschaft (DFG grant BU1803/3-1), the Alexander von Humboldt Foundation and the Freiburg Institute for Advanced Studies (FRIAS).

\bibliographystyle{apsrev4-1}
%{\bf References}
\bibliography{Surface_paper.bib}

\begin{thebibliography}{33}%
\makeatletter
\providecommand \@ifxundefined [1]{%
 \@ifx{#1\undefined}
}%
\providecommand \@ifnum [1]{%
 \ifnum #1\expandafter \@firstoftwo
 \else \expandafter \@secondoftwo
 \fi
}%
\providecommand \@ifx [1]{%
 \ifx #1\expandafter \@firstoftwo
 \else \expandafter \@secondoftwo
 \fi
}%
\providecommand \natexlab [1]{#1}%
\providecommand \enquote  [1]{``#1''}%
\providecommand \bibnamefont  [1]{#1}%
\providecommand \bibfnamefont [1]{#1}%
\providecommand \citenamefont [1]{#1}%
\providecommand \href@noop [0]{\@secondoftwo}%
\providecommand \href [0]{\begingroup \@sanitize@url \@href}%
\providecommand \@href[1]{\@@startlink{#1}\@@href}%
\providecommand \@@href[1]{\endgroup#1\@@endlink}%
\providecommand \@sanitize@url [0]{\catcode `\\12\catcode `\$12\catcode
  `\&12\catcode `\#12\catcode `\^12\catcode `\_12\catcode `\%12\relax}%
\providecommand \@@startlink[1]{}%
\providecommand \@@endlink[0]{}%
\providecommand \url  [0]{\begingroup\@sanitize@url \@url }%
\providecommand \@url [1]{\endgroup\@href {#1}{\urlprefix }}%
\providecommand \urlprefix  [0]{URL }%
\providecommand \Eprint [0]{\href }%
\providecommand \doibase [0]{http://dx.doi.org/}%
\providecommand \selectlanguage [0]{\@gobble}%
\providecommand \bibinfo  [0]{\@secondoftwo}%
\providecommand \bibfield  [0]{\@secondoftwo}%
\providecommand \translation [1]{[#1]}%
\providecommand \BibitemOpen [0]{}%
\providecommand \bibitemStop [0]{}%
\providecommand \bibitemNoStop [0]{.\EOS\space}%
\providecommand \EOS [0]{\spacefactor3000\relax}%
\providecommand \BibitemShut  [1]{\csname bibitem#1\endcsname}%
\let\auto@bib@innerbib\@empty
%</preamble>
\bibitem [{\citenamefont {Lee}\ and\ \citenamefont {Sigmund}(2001)}]{Lee2001}%
  \BibitemOpen
  \bibfield  {author} {\bibinfo {author} {\bibfnamefont {S.-W.}\ \bibnamefont
  {Lee}}\ and\ \bibinfo {author} {\bibfnamefont {W.~M.}\ \bibnamefont
  {Sigmund}},\ }\href@noop {} {\bibfield  {journal} {\bibinfo  {journal} {J.
  Colloid Interface Sci.}\ }\textbf {\bibinfo {volume} {243}},\ \bibinfo
  {pages} {365} (\bibinfo {year} {2001})}\BibitemShut {NoStop}%
\bibitem [{\citenamefont {Dill}(1990)}]{dill1990dominant}%
  \BibitemOpen
  \bibfield  {author} {\bibinfo {author} {\bibfnamefont {K.~A.}\ \bibnamefont
  {Dill}},\ }\href@noop {} {\bibfield  {journal} {\bibinfo  {journal}
  {Biochemistry}\ }\textbf {\bibinfo {volume} {29}},\ \bibinfo {pages} {7133}
  (\bibinfo {year} {1990})}\BibitemShut {NoStop}%
\bibitem [{\citenamefont {Nicholls}\ \emph {et~al.}(1991)\citenamefont
  {Nicholls}, \citenamefont {Sharp},\ and\ \citenamefont
  {Honig}}]{nicholls1991protein}%
  \BibitemOpen
  \bibfield  {author} {\bibinfo {author} {\bibfnamefont {A.}~\bibnamefont
  {Nicholls}}, \bibinfo {author} {\bibfnamefont {K.~A.}\ \bibnamefont {Sharp}},
  \ and\ \bibinfo {author} {\bibfnamefont {B.}~\bibnamefont {Honig}},\
  }\href@noop {} {\bibfield  {journal} {\bibinfo  {journal} {Proteins:
  Structure, Function, and Bioinformatics}\ }\textbf {\bibinfo {volume} {11}},\
  \bibinfo {pages} {281} (\bibinfo {year} {1991})}\BibitemShut {NoStop}%
\bibitem [{\citenamefont {Yang}\ \emph {et~al.}(2013)\citenamefont {Yang},
  \citenamefont {Adam}, \citenamefont {Nichol},\ and\ \citenamefont
  {Cockroft}}]{yang2013much}%
  \BibitemOpen
  \bibfield  {author} {\bibinfo {author} {\bibfnamefont {L.}~\bibnamefont
  {Yang}}, \bibinfo {author} {\bibfnamefont {C.}~\bibnamefont {Adam}}, \bibinfo
  {author} {\bibfnamefont {G.~S.}\ \bibnamefont {Nichol}}, \ and\ \bibinfo
  {author} {\bibfnamefont {S.~L.}\ \bibnamefont {Cockroft}},\ }\href@noop {}
  {\bibfield  {journal} {\bibinfo  {journal} {Nature chemistry}\ }\textbf
  {\bibinfo {volume} {5}},\ \bibinfo {pages} {1006} (\bibinfo {year}
  {2013})}\BibitemShut {NoStop}%
\bibitem [{\citenamefont {French}\ \emph {et~al.}(2010)\citenamefont {French},
  \citenamefont {Parsegian}, \citenamefont {Podgornik}, \citenamefont {Rajter},
  \citenamefont {Jagota}, \citenamefont {Luo}, \citenamefont {Asthagiri},
  \citenamefont {Chaudhury}, \citenamefont {Chiang}, \citenamefont {Granick},
  \citenamefont {Kalinin}, \citenamefont {Kardar}, \citenamefont {Kjellander},
  \citenamefont {Langreth}, \citenamefont {Lewis}, \citenamefont {Lustig},
  \citenamefont {Wesolowski}, \citenamefont {Wettlaufer}, \citenamefont
  {Ching}, \citenamefont {Finnis}, \citenamefont {Houlihan}, \citenamefont {von
  Lilienfeld}, \citenamefont {van Oss},\ and\ \citenamefont
  {Zemb}}]{RevModPhys.82.1887}%
  \BibitemOpen
  \bibfield  {author} {\bibinfo {author} {\bibfnamefont {R.~H.}\ \bibnamefont
  {French}}, \bibinfo {author} {\bibfnamefont {V.~A.}\ \bibnamefont
  {Parsegian}}, \bibinfo {author} {\bibfnamefont {R.}~\bibnamefont
  {Podgornik}}, \bibinfo {author} {\bibfnamefont {R.~F.}\ \bibnamefont
  {Rajter}}, \bibinfo {author} {\bibfnamefont {A.}~\bibnamefont {Jagota}},
  \bibinfo {author} {\bibfnamefont {J.}~\bibnamefont {Luo}}, \bibinfo {author}
  {\bibfnamefont {D.}~\bibnamefont {Asthagiri}}, \bibinfo {author}
  {\bibfnamefont {M.~K.}\ \bibnamefont {Chaudhury}}, \bibinfo {author}
  {\bibfnamefont {Y.-m.}\ \bibnamefont {Chiang}}, \bibinfo {author}
  {\bibfnamefont {S.}~\bibnamefont {Granick}}, \bibinfo {author} {\bibfnamefont
  {S.}~\bibnamefont {Kalinin}}, \bibinfo {author} {\bibfnamefont
  {M.}~\bibnamefont {Kardar}}, \bibinfo {author} {\bibfnamefont
  {R.}~\bibnamefont {Kjellander}}, \bibinfo {author} {\bibfnamefont {D.~C.}\
  \bibnamefont {Langreth}}, \bibinfo {author} {\bibfnamefont {J.}~\bibnamefont
  {Lewis}}, \bibinfo {author} {\bibfnamefont {S.}~\bibnamefont {Lustig}},
  \bibinfo {author} {\bibfnamefont {D.}~\bibnamefont {Wesolowski}}, \bibinfo
  {author} {\bibfnamefont {J.~S.}\ \bibnamefont {Wettlaufer}}, \bibinfo
  {author} {\bibfnamefont {W.-Y.}\ \bibnamefont {Ching}}, \bibinfo {author}
  {\bibfnamefont {M.}~\bibnamefont {Finnis}}, \bibinfo {author} {\bibfnamefont
  {F.}~\bibnamefont {Houlihan}}, \bibinfo {author} {\bibfnamefont {O.~A.}\
  \bibnamefont {von Lilienfeld}}, \bibinfo {author} {\bibfnamefont {C.~J.}\
  \bibnamefont {van Oss}}, \ and\ \bibinfo {author} {\bibfnamefont
  {T.}~\bibnamefont {Zemb}},\ }\href {\doibase 10.1103/RevModPhys.82.1887}
  {\bibfield  {journal} {\bibinfo  {journal} {Rev. Mod. Phys.}\ }\textbf
  {\bibinfo {volume} {82}},\ \bibinfo {pages} {1887} (\bibinfo {year}
  {2010})}\BibitemShut {NoStop}%
\bibitem [{\citenamefont {Geraci}\ \emph {et~al.}(2010)\citenamefont {Geraci},
  \citenamefont {Papp},\ and\ \citenamefont
  {Kitching}}]{Geraci2010Short-rangeMicrospheres}%
  \BibitemOpen
  \bibfield  {author} {\bibinfo {author} {\bibfnamefont {A.~A.}\ \bibnamefont
  {Geraci}}, \bibinfo {author} {\bibfnamefont {S.~B.}\ \bibnamefont {Papp}}, \
  and\ \bibinfo {author} {\bibfnamefont {J.}~\bibnamefont {Kitching}},\
  }\href@noop {} {\bibfield  {journal} {\bibinfo  {journal} {Phys. Rev. Lett.}\
  }\textbf {\bibinfo {volume} {105}},\ \bibinfo {pages} {101101} (\bibinfo
  {year} {2010})}\BibitemShut {NoStop}%
\bibitem [{\citenamefont {Kapner}\ \emph {et~al.}(2007)\citenamefont {Kapner},
  \citenamefont {Cook}, \citenamefont {Adelberger}, \citenamefont {Gundlach},
  \citenamefont {Heckel}, \citenamefont {Hoyle},\ and\ \citenamefont
  {Swanson}}]{Kapner2007}%
  \BibitemOpen
  \bibfield  {author} {\bibinfo {author} {\bibfnamefont {D.~J.}\ \bibnamefont
  {Kapner}}, \bibinfo {author} {\bibfnamefont {T.~S.}\ \bibnamefont {Cook}},
  \bibinfo {author} {\bibfnamefont {E.~G.}\ \bibnamefont {Adelberger}},
  \bibinfo {author} {\bibfnamefont {J.~H.}\ \bibnamefont {Gundlach}}, \bibinfo
  {author} {\bibfnamefont {B.~R.}\ \bibnamefont {Heckel}}, \bibinfo {author}
  {\bibfnamefont {C.~D.}\ \bibnamefont {Hoyle}}, \ and\ \bibinfo {author}
  {\bibfnamefont {H.~E.}\ \bibnamefont {Swanson}},\ }\href@noop {} {\bibfield
  {journal} {\bibinfo  {journal} {Phys. Rev. Lett.}\ }\textbf {\bibinfo
  {volume} {98}},\ \bibinfo {pages} {021101} (\bibinfo {year}
  {2007})}\BibitemShut {NoStop}%
\bibitem [{\citenamefont {Moore}\ \emph {et~al.}(2014)\citenamefont {Moore},
  \citenamefont {Rider},\ and\ \citenamefont {Gratta}}]{Moore2014}%
  \BibitemOpen
  \bibfield  {author} {\bibinfo {author} {\bibfnamefont {D.~C.}\ \bibnamefont
  {Moore}}, \bibinfo {author} {\bibfnamefont {A.~D.}\ \bibnamefont {Rider}}, \
  and\ \bibinfo {author} {\bibfnamefont {G.}~\bibnamefont {Gratta}},\
  }\href@noop {} {\bibfield  {journal} {\bibinfo  {journal} {Phys. Rev. Lett.}\
  }\textbf {\bibinfo {volume} {113}},\ \bibinfo {pages} {251801} (\bibinfo
  {year} {2014})}\BibitemShut {NoStop}%
\bibitem [{\citenamefont {Schm{\"o}le}\ \emph {et~al.}(2016)\citenamefont
  {Schm{\"o}le}, \citenamefont {Dragosits}, \citenamefont {Hepach},\ and\
  \citenamefont {Aspelmeyer}}]{schmole2016micromechanical}%
  \BibitemOpen
  \bibfield  {author} {\bibinfo {author} {\bibfnamefont {J.}~\bibnamefont
  {Schm{\"o}le}}, \bibinfo {author} {\bibfnamefont {M.}~\bibnamefont
  {Dragosits}}, \bibinfo {author} {\bibfnamefont {H.}~\bibnamefont {Hepach}}, \
  and\ \bibinfo {author} {\bibfnamefont {M.}~\bibnamefont {Aspelmeyer}},\
  }\href@noop {} {\bibfield  {journal} {\bibinfo  {journal} {Classical and
  Quantum Gravity}\ }\textbf {\bibinfo {volume} {33}},\ \bibinfo {pages}
  {125031} (\bibinfo {year} {2016})}\BibitemShut {NoStop}%
\bibitem [{\citenamefont {Hamilton}\ \emph {et~al.}(2015)\citenamefont
  {Hamilton}, \citenamefont {Jaffe}, \citenamefont {Haslinger}, \citenamefont
  {Simmons}, \citenamefont {M{\"u}ller},\ and\ \citenamefont
  {Khoury}}]{hamilton2015atom}%
  \BibitemOpen
  \bibfield  {author} {\bibinfo {author} {\bibfnamefont {P.}~\bibnamefont
  {Hamilton}}, \bibinfo {author} {\bibfnamefont {M.}~\bibnamefont {Jaffe}},
  \bibinfo {author} {\bibfnamefont {P.}~\bibnamefont {Haslinger}}, \bibinfo
  {author} {\bibfnamefont {Q.}~\bibnamefont {Simmons}}, \bibinfo {author}
  {\bibfnamefont {H.}~\bibnamefont {M{\"u}ller}}, \ and\ \bibinfo {author}
  {\bibfnamefont {J.}~\bibnamefont {Khoury}},\ }\href@noop {} {\bibfield
  {journal} {\bibinfo  {journal} {Science}\ }\textbf {\bibinfo {volume}
  {349}},\ \bibinfo {pages} {849} (\bibinfo {year} {2015})}\BibitemShut
  {NoStop}%
\bibitem [{\citenamefont {Shih}\ and\ \citenamefont
  {Parsegian}(1975)}]{Shih1975}%
  \BibitemOpen
  \bibfield  {author} {\bibinfo {author} {\bibfnamefont {A.}~\bibnamefont
  {Shih}}\ and\ \bibinfo {author} {\bibfnamefont {V.~A.}\ \bibnamefont
  {Parsegian}},\ }\href@noop {} {\bibfield  {journal} {\bibinfo  {journal}
  {Phys. Rev. A}\ }\textbf {\bibinfo {volume} {12}},\ \bibinfo {pages} {835}
  (\bibinfo {year} {1975})}\BibitemShut {NoStop}%
\bibitem [{\citenamefont {Sukenik}\ \emph {et~al.}(1993)\citenamefont
  {Sukenik}, \citenamefont {Boshier}, \citenamefont {Cho}, \citenamefont
  {Sandoghdar},\ and\ \citenamefont {Hinds}}]{Sukenik1993}%
  \BibitemOpen
  \bibfield  {author} {\bibinfo {author} {\bibfnamefont {C.~I.}\ \bibnamefont
  {Sukenik}}, \bibinfo {author} {\bibfnamefont {M.~G.}\ \bibnamefont
  {Boshier}}, \bibinfo {author} {\bibfnamefont {D.}~\bibnamefont {Cho}},
  \bibinfo {author} {\bibfnamefont {V.}~\bibnamefont {Sandoghdar}}, \ and\
  \bibinfo {author} {\bibfnamefont {E.~A.}\ \bibnamefont {Hinds}},\ }\href@noop
  {} {\bibfield  {journal} {\bibinfo  {journal} {Phys. Rev. Lett.}\ }\textbf
  {\bibinfo {volume} {70}},\ \bibinfo {pages} {560} (\bibinfo {year}
  {1993})}\BibitemShut {NoStop}%
\bibitem [{\citenamefont {Obrecht}\ \emph {et~al.}(2007)\citenamefont
  {Obrecht}, \citenamefont {Wild}, \citenamefont {Antezza}, \citenamefont
  {Pitaevskii}, \citenamefont {Stringari},\ and\ \citenamefont
  {Cornell}}]{Obrecht2007}%
  \BibitemOpen
  \bibfield  {author} {\bibinfo {author} {\bibfnamefont {J.~M.}\ \bibnamefont
  {Obrecht}}, \bibinfo {author} {\bibfnamefont {R.~J.}\ \bibnamefont {Wild}},
  \bibinfo {author} {\bibfnamefont {M.}~\bibnamefont {Antezza}}, \bibinfo
  {author} {\bibfnamefont {L.~P.}\ \bibnamefont {Pitaevskii}}, \bibinfo
  {author} {\bibfnamefont {S.}~\bibnamefont {Stringari}}, \ and\ \bibinfo
  {author} {\bibfnamefont {E.~A.}\ \bibnamefont {Cornell}},\ }\href@noop {}
  {\bibfield  {journal} {\bibinfo  {journal} {Phys. Rev. Lett.}\ }\textbf
  {\bibinfo {volume} {98}},\ \bibinfo {pages} {063201} (\bibinfo {year}
  {2007})}\BibitemShut {NoStop}%
\bibitem [{\citenamefont {Bender}\ \emph {et~al.}(2014)\citenamefont {Bender},
  \citenamefont {Stehle}, \citenamefont {Zimmermann}, \citenamefont {Slama},
  \citenamefont {Fiedler}, \citenamefont {Scheel}, \citenamefont {Buhmann},\
  and\ \citenamefont {Marachevsky}}]{bender2014}%
  \BibitemOpen
  \bibfield  {author} {\bibinfo {author} {\bibfnamefont {H.}~\bibnamefont
  {Bender}}, \bibinfo {author} {\bibfnamefont {C.}~\bibnamefont {Stehle}},
  \bibinfo {author} {\bibfnamefont {C.}~\bibnamefont {Zimmermann}}, \bibinfo
  {author} {\bibfnamefont {S.}~\bibnamefont {Slama}}, \bibinfo {author}
  {\bibfnamefont {J.}~\bibnamefont {Fiedler}}, \bibinfo {author} {\bibfnamefont
  {S.}~\bibnamefont {Scheel}}, \bibinfo {author} {\bibfnamefont {S.~Y.}\
  \bibnamefont {Buhmann}}, \ and\ \bibinfo {author} {\bibfnamefont {V.~N.}\
  \bibnamefont {Marachevsky}},\ }\href {\doibase 10.1103/PhysRevX.4.011029}
  {\bibfield  {journal} {\bibinfo  {journal} {Phys. Rev. X}\ }\textbf {\bibinfo
  {volume} {4}},\ \bibinfo {pages} {011029} (\bibinfo {year} {2014})},\ \Eprint
  {http://arxiv.org/abs/1305.1832} {arXiv:1305.1832} \BibitemShut {NoStop}%
\bibitem [{\citenamefont {Lamoreaux}(1997)}]{Lamoreaux1997}%
  \BibitemOpen
  \bibfield  {author} {\bibinfo {author} {\bibfnamefont {S.~K.}\ \bibnamefont
  {Lamoreaux}},\ }\href {\doibase 10.1103/PhysRevLett.78.5} {\bibfield
  {journal} {\bibinfo  {journal} {Phys. Rev. Lett.}\ }\textbf {\bibinfo
  {volume} {78}},\ \bibinfo {pages} {5} (\bibinfo {year} {1997})}\BibitemShut
  {NoStop}%
\bibitem [{\citenamefont {Mohideen}\ and\ \citenamefont
  {Roy}()}]{Mohideen1998}%
  \BibitemOpen
  \bibfield  {author} {\bibinfo {author} {\bibfnamefont {U.}~\bibnamefont
  {Mohideen}}\ and\ \bibinfo {author} {\bibfnamefont {A.}~\bibnamefont {Roy}},\
  }\href@noop {} {\bibinfo  {journal} {Phys. Rev. Lett.}\ ,\ \bibinfo {pages}
  {4549}}\BibitemShut {NoStop}%
\bibitem [{\citenamefont {Binnig}\ \emph {et~al.}(1986)\citenamefont {Binnig},
  \citenamefont {Quate},\ and\ \citenamefont {Gerber}}]{PhysRevLett.56.930}%
  \BibitemOpen
\bibfield  {journal} {  }\bibfield  {author} {\bibinfo {author} {\bibfnamefont
  {G.}~\bibnamefont {Binnig}}, \bibinfo {author} {\bibfnamefont {C.~F.}\
  \bibnamefont {Quate}}, \ and\ \bibinfo {author} {\bibfnamefont
  {C.}~\bibnamefont {Gerber}},\ }\href {\doibase 10.1103/PhysRevLett.56.930}
  {\bibfield  {journal} {\bibinfo  {journal} {Phys. Rev. Lett.}\ }\textbf
  {\bibinfo {volume} {56}},\ \bibinfo {pages} {930} (\bibinfo {year}
  {1986})}\BibitemShut {NoStop}%
\bibitem [{\citenamefont {Casimir}(1948)}]{CasimirOriginal1948}%
  \BibitemOpen
  \bibfield  {author} {\bibinfo {author} {\bibfnamefont {H.}~\bibnamefont
  {Casimir}},\ }\href {\doibase citeulike-article-id:8810715} {\bibfield
  {journal} {\bibinfo  {journal} {Proc. K. Ned. Akad.}\ }\textbf {\bibinfo
  {volume} {360}},\ \bibinfo {pages} {793} (\bibinfo {year}
  {1948})}\BibitemShut {NoStop}%
\bibitem [{\citenamefont {Casimir}\ and\ \citenamefont
  {Polder}(1948)}]{Casimir1948}%
  \BibitemOpen
  \bibfield  {author} {\bibinfo {author} {\bibfnamefont {H.~B.~G.}\
  \bibnamefont {Casimir}}\ and\ \bibinfo {author} {\bibfnamefont
  {D.}~\bibnamefont {Polder}},\ }\href@noop {} {\bibfield  {journal} {\bibinfo
  {journal} {Phys. Rev.}\ }\textbf {\bibinfo {volume} {73}},\ \bibinfo {pages}
  {360} (\bibinfo {year} {1948})}\BibitemShut {NoStop}%
\bibitem [{\citenamefont {Rashid}\ \emph {et~al.}(2016)\citenamefont {Rashid},
  \citenamefont {Tufarelli}, \citenamefont {Bateman}, \citenamefont {Vovrosh},
  \citenamefont {Hempston}, \citenamefont {Kim},\ and\ \citenamefont
  {Ulbricht}}]{Rashid2016}%
  \BibitemOpen
  \bibfield  {author} {\bibinfo {author} {\bibfnamefont {M.}~\bibnamefont
  {Rashid}}, \bibinfo {author} {\bibfnamefont {T.}~\bibnamefont {Tufarelli}},
  \bibinfo {author} {\bibfnamefont {J.}~\bibnamefont {Bateman}}, \bibinfo
  {author} {\bibfnamefont {J.}~\bibnamefont {Vovrosh}}, \bibinfo {author}
  {\bibfnamefont {D.}~\bibnamefont {Hempston}}, \bibinfo {author}
  {\bibfnamefont {M.~S.}\ \bibnamefont {Kim}}, \ and\ \bibinfo {author}
  {\bibfnamefont {H.}~\bibnamefont {Ulbricht}},\ }\href@noop {} {\bibfield
  {journal} {\bibinfo  {journal} {Phys. Rev. Lett.}\ }\textbf {\bibinfo
  {volume} {117}},\ \bibinfo {pages} {273601} (\bibinfo {year}
  {2016})}\BibitemShut {NoStop}%
\bibitem [{\citenamefont {Palik}(1985)}]{Palik1985}%
  \BibitemOpen
  \bibfield  {author} {\bibinfo {author} {\bibfnamefont {E.~D.}\ \bibnamefont
  {Palik}},\ }\href@noop {} {\emph {\bibinfo {title} {{Handbook of optical
  constants of solids}}}}\ (\bibinfo  {publisher} {Academic Press},\ \bibinfo
  {year} {1985})\BibitemShut {NoStop}%
\bibitem [{\citenamefont {Spruch}\ and\ \citenamefont
  {Tikochinsky}(1993)}]{Spruch1993}%
  \BibitemOpen
  \bibfield  {author} {\bibinfo {author} {\bibfnamefont {L.}~\bibnamefont
  {Spruch}}\ and\ \bibinfo {author} {\bibfnamefont {Y.}~\bibnamefont
  {Tikochinsky}},\ }\href@noop {} {\bibfield  {journal} {\bibinfo  {journal}
  {Phys. Rev. A}\ }\textbf {\bibinfo {volume} {48}},\ \bibinfo {pages} {4213}
  (\bibinfo {year} {1993})}\BibitemShut {NoStop}%
\bibitem [{\citenamefont {Rondin}\ \emph {et~al.}(2017)\citenamefont {Rondin},
  \citenamefont {Gieseler}, \citenamefont {Ricci}, \citenamefont {Quidant},
  \citenamefont {Dellago},\ and\ \citenamefont {Novotny}}]{Rondin2017}%
  \BibitemOpen
  \bibfield  {author} {\bibinfo {author} {\bibfnamefont {L.}~\bibnamefont
  {Rondin}}, \bibinfo {author} {\bibfnamefont {J.}~\bibnamefont {Gieseler}},
  \bibinfo {author} {\bibfnamefont {F.}~\bibnamefont {Ricci}}, \bibinfo
  {author} {\bibfnamefont {R.}~\bibnamefont {Quidant}}, \bibinfo {author}
  {\bibfnamefont {C.}~\bibnamefont {Dellago}}, \ and\ \bibinfo {author}
  {\bibfnamefont {L.}~\bibnamefont {Novotny}},\ }\href@noop {} {\bibfield
  {journal} {\bibinfo  {journal} {Nature Nanotechnology}\ ,\ \bibinfo {pages}
  {doi:10.1038/nnano.2017.198}} (\bibinfo {year} {2017})}\BibitemShut {NoStop}%
\bibitem [{\citenamefont {Frimmer}\ \emph {et~al.}(2017)\citenamefont
  {Frimmer}, \citenamefont {Luszcz}, \citenamefont {Ferreiro}, \citenamefont
  {Jain}, \citenamefont {Hebestreit},\ and\ \citenamefont
  {Novotny}}]{Frimmer2017}%
  \BibitemOpen
  \bibfield  {author} {\bibinfo {author} {\bibfnamefont {M.}~\bibnamefont
  {Frimmer}}, \bibinfo {author} {\bibfnamefont {K.}~\bibnamefont {Luszcz}},
  \bibinfo {author} {\bibfnamefont {S.}~\bibnamefont {Ferreiro}}, \bibinfo
  {author} {\bibfnamefont {V.}~\bibnamefont {Jain}}, \bibinfo {author}
  {\bibfnamefont {E.}~\bibnamefont {Hebestreit}}, \ and\ \bibinfo {author}
  {\bibfnamefont {L.}~\bibnamefont {Novotny}},\ }\href@noop {} {\bibfield
  {journal} {\bibinfo  {journal} {Phys. Rev. A}\ }\textbf {\bibinfo {volume}
  {95}},\ \bibinfo {pages} {061801} (\bibinfo {year} {2017})}\BibitemShut
  {NoStop}%
\bibitem [{\citenamefont {Hempston}\ \emph {et~al.}(2017)\citenamefont
  {Hempston}, \citenamefont {Vovrosh}, \citenamefont {Toro{\v{s}}},
  \citenamefont {Winstone}, \citenamefont {Rashid},\ and\ \citenamefont
  {Ulbricht}}]{hempston2017force}%
  \BibitemOpen
  \bibfield  {author} {\bibinfo {author} {\bibfnamefont {D.}~\bibnamefont
  {Hempston}}, \bibinfo {author} {\bibfnamefont {J.}~\bibnamefont {Vovrosh}},
  \bibinfo {author} {\bibfnamefont {M.}~\bibnamefont {Toro{\v{s}}}}, \bibinfo
  {author} {\bibfnamefont {G.}~\bibnamefont {Winstone}}, \bibinfo {author}
  {\bibfnamefont {M.}~\bibnamefont {Rashid}}, \ and\ \bibinfo {author}
  {\bibfnamefont {H.}~\bibnamefont {Ulbricht}},\ }\href@noop {} {\bibfield
  {journal} {\bibinfo  {journal} {Applied Physics Letters}\ }\textbf {\bibinfo
  {volume} {111}},\ \bibinfo {pages} {133111} (\bibinfo {year}
  {2017})}\BibitemShut {NoStop}%
\bibitem [{\citenamefont {Sushkov}\ \emph {et~al.}(2011)\citenamefont
  {Sushkov}, \citenamefont {Kim}, \citenamefont {Dalvit},\ and\ \citenamefont
  {Lamoreaux}}]{sushkov2011observation}%
  \BibitemOpen
  \bibfield  {author} {\bibinfo {author} {\bibfnamefont {A.}~\bibnamefont
  {Sushkov}}, \bibinfo {author} {\bibfnamefont {W.}~\bibnamefont {Kim}},
  \bibinfo {author} {\bibfnamefont {D.}~\bibnamefont {Dalvit}}, \ and\ \bibinfo
  {author} {\bibfnamefont {S.}~\bibnamefont {Lamoreaux}},\ }\href@noop {}
  {\bibfield  {journal} {\bibinfo  {journal} {Nature Physics}\ }\textbf
  {\bibinfo {volume} {7}},\ \bibinfo {pages} {230} (\bibinfo {year}
  {2011})}\BibitemShut {NoStop}%
\bibitem [{\citenamefont {Behunin}\ \emph {et~al.}(2012)\citenamefont
  {Behunin}, \citenamefont {Zeng}, \citenamefont {Dalvit},\ and\ \citenamefont
  {Reynaud}}]{behunin2012electrostatic}%
  \BibitemOpen
  \bibfield  {author} {\bibinfo {author} {\bibfnamefont {R.}~\bibnamefont
  {Behunin}}, \bibinfo {author} {\bibfnamefont {Y.}~\bibnamefont {Zeng}},
  \bibinfo {author} {\bibfnamefont {D.}~\bibnamefont {Dalvit}}, \ and\ \bibinfo
  {author} {\bibfnamefont {S.}~\bibnamefont {Reynaud}},\ }\href@noop {}
  {\bibfield  {journal} {\bibinfo  {journal} {Physical Review A}\ }\textbf
  {\bibinfo {volume} {86}},\ \bibinfo {pages} {052509} (\bibinfo {year}
  {2012})}\BibitemShut {NoStop}%
\bibitem [{\citenamefont {Vovrosh}\ \emph {et~al.}(2017)\citenamefont
  {Vovrosh}, \citenamefont {Rashid}, \citenamefont {Hempston}, \citenamefont
  {Bateman}, \citenamefont {Paternostro},\ and\ \citenamefont
  {Ulbricht}}]{vovrosh2017parametric}%
  \BibitemOpen
  \bibfield  {author} {\bibinfo {author} {\bibfnamefont {J.}~\bibnamefont
  {Vovrosh}}, \bibinfo {author} {\bibfnamefont {M.}~\bibnamefont {Rashid}},
  \bibinfo {author} {\bibfnamefont {D.}~\bibnamefont {Hempston}}, \bibinfo
  {author} {\bibfnamefont {J.}~\bibnamefont {Bateman}}, \bibinfo {author}
  {\bibfnamefont {M.}~\bibnamefont {Paternostro}}, \ and\ \bibinfo {author}
  {\bibfnamefont {H.}~\bibnamefont {Ulbricht}},\ }\href@noop {} {\bibfield
  {journal} {\bibinfo  {journal} {JOSA B}\ }\textbf {\bibinfo {volume} {34}},\
  \bibinfo {pages} {1421} (\bibinfo {year} {2017})}\BibitemShut {NoStop}%
\bibitem [{\citenamefont {Derevianko}\ \emph {et~al.}(1999)\citenamefont
  {Derevianko}, \citenamefont {Johnson}, \citenamefont {Safronova},\ and\
  \citenamefont {Babb}}]{Derevianko1999}%
  \BibitemOpen
  \bibfield  {author} {\bibinfo {author} {\bibfnamefont {A.}~\bibnamefont
  {Derevianko}}, \bibinfo {author} {\bibfnamefont {W.~R.}\ \bibnamefont
  {Johnson}}, \bibinfo {author} {\bibfnamefont {M.~S.}\ \bibnamefont
  {Safronova}}, \ and\ \bibinfo {author} {\bibfnamefont {J.~F.}\ \bibnamefont
  {Babb}},\ }\href {\doibase 10.1103/PhysRevLett.82.3589} {\bibfield  {journal}
  {\bibinfo  {journal} {Phys. Rev. Lett.}\ }\textbf {\bibinfo {volume} {82}},\
  \bibinfo {pages} {3589} (\bibinfo {year} {1999})}\BibitemShut {NoStop}%
\bibitem [{\citenamefont {Manjavacas}\ \emph {et~al.}(2017)\citenamefont
  {Manjavacas}, \citenamefont {Rodr\'{\i}guez-Fortu\~no}, \citenamefont
  {Garc\'{\i}a~de Abajo},\ and\ \citenamefont {Zayats}}]{Manjavacas2017}%
  \BibitemOpen
  \bibfield  {author} {\bibinfo {author} {\bibfnamefont {A.}~\bibnamefont
  {Manjavacas}}, \bibinfo {author} {\bibfnamefont {F.~J.}\ \bibnamefont
  {Rodr\'{\i}guez-Fortu\~no}}, \bibinfo {author} {\bibfnamefont {F.~J.}\
  \bibnamefont {Garc\'{\i}a~de Abajo}}, \ and\ \bibinfo {author} {\bibfnamefont
  {A.~V.}\ \bibnamefont {Zayats}},\ }\href@noop {} {\bibfield  {journal}
  {\bibinfo  {journal} {Phys. Rev. Lett.}\ }\textbf {\bibinfo {volume} {118}},\
  \bibinfo {pages} {133605} (\bibinfo {year} {2017})}\BibitemShut {NoStop}%
\bibitem [{\citenamefont {Juan}\ \emph {et~al.}(2009)\citenamefont {Juan},
  \citenamefont {Gordon}, \citenamefont {Pang}, \citenamefont {Eftekhari},\
  and\ \citenamefont {Quidant}}]{juan2009self}%
  \BibitemOpen
  \bibfield  {author} {\bibinfo {author} {\bibfnamefont {M.~L.}\ \bibnamefont
  {Juan}}, \bibinfo {author} {\bibfnamefont {R.}~\bibnamefont {Gordon}},
  \bibinfo {author} {\bibfnamefont {Y.}~\bibnamefont {Pang}}, \bibinfo {author}
  {\bibfnamefont {F.}~\bibnamefont {Eftekhari}}, \ and\ \bibinfo {author}
  {\bibfnamefont {R.}~\bibnamefont {Quidant}},\ }\href@noop {} {\bibfield
  {journal} {\bibinfo  {journal} {Nature Physics}\ }\textbf {\bibinfo {volume}
  {5}},\ \bibinfo {pages} {915} (\bibinfo {year} {2009})}\BibitemShut {NoStop}%
\bibitem [{\citenamefont {Steinlechner}\ \emph {et~al.}(2013)\citenamefont
  {Steinlechner}, \citenamefont {Kr\"uger}, \citenamefont {Lastzka},
  \citenamefont {Steinlechner}, \citenamefont {Khalaidovski},\ and\
  \citenamefont {Schnabel}}]{Steinlechner2013}%
  \BibitemOpen
  \bibfield  {author} {\bibinfo {author} {\bibfnamefont {J.}~\bibnamefont
  {Steinlechner}}, \bibinfo {author} {\bibfnamefont {C.}~\bibnamefont
  {Kr\"uger}}, \bibinfo {author} {\bibfnamefont {N.}~\bibnamefont {Lastzka}},
  \bibinfo {author} {\bibfnamefont {S.}~\bibnamefont {Steinlechner}}, \bibinfo
  {author} {\bibfnamefont {A.}~\bibnamefont {Khalaidovski}}, \ and\ \bibinfo
  {author} {\bibfnamefont {R.}~\bibnamefont {Schnabel}},\ }\href@noop {}
  {\bibfield  {journal} {\bibinfo  {journal} {Classical Quantum Gravity}\
  }\textbf {\bibinfo {volume} {30}},\ \bibinfo {pages} {095007} (\bibinfo
  {year} {2013})}\BibitemShut {NoStop}%
\bibitem [{\citenamefont {Dzyaloshinskii}\ \emph {et~al.}(1961)\citenamefont
  {Dzyaloshinskii}, \citenamefont {Lifshitz},\ and\ \citenamefont
  {Pitaevskii}}]{Dzyaloshinskii1961}%
  \BibitemOpen
  \bibfield  {author} {\bibinfo {author} {\bibfnamefont {I.}~\bibnamefont
  {Dzyaloshinskii}}, \bibinfo {author} {\bibfnamefont {E.}~\bibnamefont
  {Lifshitz}}, \ and\ \bibinfo {author} {\bibfnamefont {L.}~\bibnamefont
  {Pitaevskii}},\ }\href {\doibase 10.1080/00018736100101281} {\bibfield
  {journal} {\bibinfo  {journal} {Adv. Phys.}\ }\textbf {\bibinfo {volume}
  {10}},\ \bibinfo {pages} {165} (\bibinfo {year} {1961})}\BibitemShut
  {NoStop}%
\end{thebibliography}%

%{\bf Author contributions}

%G.W. and M.R. performed the experiments and G.W. analysed the data. R.B. and S.B. developed the theory. H.U. initiated and supervised the project. All the authors discussed
%the results, the comparison of experiment with theory, the implications and the figures, and wrote the manuscript.

%{\bf Additional information}

%Supplementary information is available in the online version of the paper. Reprints and
%permissions information is available online at www.nature.com/reprints. Publisher's note:
%Springer Nature remains neutral with regard to jurisdictional claims in published maps and
%institutional affiliations. Correspondence and requests for materials should be addressed to H.U.

%{\bf Competing financial interests}

%The authors declare no competing financial interests

\clearpage

{\bf Methods}

{\bf Experimental set-up.} We optically trap a dielectric SiO$_2$ nanosphere in the focus of a high numerical aperture (NA=0.9) parabolic mirror. The trapping laser wavelengths and power are 1550 nm and 1 W, respectively. All experiments are conducted at a pressure of $10^{-2}$ mbar while no active cooling of the centre of mass motion of the particle has been implemented, in order to allow for large oscillation amplitudes of the particle in the trap. The motion of the particle is still both damped and driven by collisions with background gas. However the effect of the interaction of the particle with the surface already becomes clearly visible and is measurable as it affects the shape of the trap. The experimental setup is shown in figure  \ref{fig:setup} and more details about optical trapping using parabolic mirrors can be found elsewhere \cite{Rashid2016, vovrosh2017parametric}. 
In some more detail, a three axis micrometer stage is used to vary the distance between a 200 $\mu$m thick, highly n-doped and double-sided polished silicon planar surface (Si wafer), which is transparent at 1550 nm \cite{Steinlechner2013} and the optically defined nanoparticle trapping site. The surface is moved in discrete intervals, decreasing the distance to the levitated particle. At each stage position the nanoparticle's motion is recorded by a homodyne detection scheme with high spatial resolution utilized in previous studies \cite{Rashid2016, vovrosh2017parametric}. At each stage position, the oscillation of the particle explores a region of several hundred nanometers, with the exact distance being determined by the potential stiffness. This allows us to reconstruct the overall surface potential in piece-wise steps.

{\bf Equation of motion without surface force.} For pressures below approximately 10 mbar, the motion of the particle can be treated as three decoupled one-dimensional driven damped harmonic oscillators, each described by an equation of motion of the form: $\ddot{y} + \gamma \dot{y} + \Omega y = F(t)/m$, with $\gamma$ describing the damping of the motional degree of freedom of the particle, $\Omega$ is the natural frequency of that oscillator, $m$ represents its mass, $y$ the oscillator's displacement and $F(t)$ describes the fluctuating forces acting on the particle according to random collisions with background gas particles.

\begin{figure}
\centering
\includegraphics[width=\columnwidth]{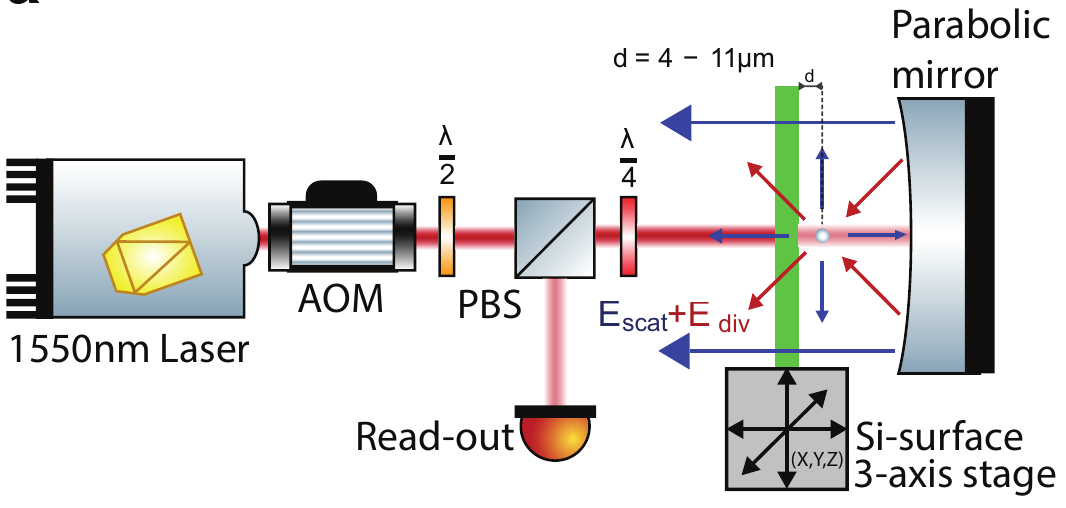}
\caption{{\bf Experimental setup.}  The 1550 nm light is focused by reflection off a parabolic mirror to a spot of waist of about 1$\mu$m. The light wave Rayleigh scattered by the particle is superposed with a diverging reference wave, which gives the high position resolution, 1pm, of the the detection. The light is detected by a cooled InGaAs photodiode. The silicon surface, which is optically transparent at the trapping laser wavelength, is mounted on a vacuum-compatible $x,y,z$-stage in order to control the particle-surface distance. 
}
\label{fig:setup}
\end{figure}

{\bf Competing optical effects.}  In principle, it seems possible that the emergence of the anharmonic trapping potential is due to an optical effect rather than being induced by dispersive surface interactions. A tiny fraction of light incident to the particle and surface is reflected by the surface in such a way that it reduces the laser power incident to the trap and therefore lowers the $x^2$ trapping potential. This effect is however separable from the effect of the surface interactions as those scale as $\frac{1}{x^n}$ and therefore are always asymmetric, while the optical potential is composed of symmetric $x^2$ terms for the region of the trap explored by the particle. 
Further when the purely optical potential dynamics is parametrically driven into its non-linear regions as for instance for large oscillation amplitudes, $x^4$ and higher order symmetric Duffing terms appear. This means anharmonic terms, which would result in a non-symmetric potential cannot be generated for a static optical potential in this geometry. Thus the anharmonic potential we observe cannot be explained by an optical effect.

We experimentally check the laser power dependence of the trap potential and observe that the potential as well as the spring functions remain of $x^2$ and linear type, respectively (see figure \ref{fig:springs_and_potentials} in supplement). This further reinforces that the observed anharmonicity of the trapping potential for the particle close to the surface is not of optical origin.

Another effect maybe caused by the multiple reflection of light between surface and particle, known as optical binding, which have a similar scaling with distance as some dispersion forces. We argue that such multiple reflections will change the laser power forming the trap and conclude for the same rationale given above optical binding cannot explain the observed anharmonicity in these experiments.

%%%%%%%%%%Supplement%%%%%%%%%%%%%%%

\clearpage
\onecolumngrid

\section{Supplementary Material for: Direct measurement of short-range forces with a levitated nanoparticle} 
Here we give more details on the extraction of the mass and electrical charge of the particle, the extraction of the trapping potential form measured time-domain data, and the energy level extraction for the anharmonic Morse-like potential.

\subsection{Mass/Radius of the trapped particle}
To extract the mass of the particle from the experimental data, we adopt a method from Rondin et al. \cite{Rondin2017}, but in one dimension rather than three. The mass is computed by comparing two potentials, namely the steady state potential $U_\text{stst}$ and the kinematic potential $U_\text{kin}$, with the mass $m$ as the only free parameter.  The potential $U_\text{stst}$ computed from the steady state solution of a Langevin equation of a particle in thermodynamic equilibrium with a random background field undergoing a random walk in a potential giving, 
\begin{equation}
U_\text{stst} = \frac{k_B T}{\gamma} \ln(\rho(x)), 
\label{steadystatepotentialequation}
\end{equation}
where $\gamma$ is the damping constant of the damped harmonic oscillator describing the motion of the particle in the trap. The dominant contribution to the damping comes from random kicks of background gas particles with the trapped particle at thermal equilibrium with the environment at 300K. $\rho(x)$ represents the position distribution of the trapped particle. Potential $U_\text{kin}$ is extracted experimentally from the time trace of the motion of the particle in the trap according to
\begin{equation}
U_\text{kin} = \int F(t) \cdot dx + c ,
\label{kinequation}
\end{equation}
where $c$ is a constant of integration, and $F(t)$ is the total time-dependent force acting on the particle.  $F(t)$  is computed by taking the finite difference with respect to the particle's time trace. 
\begin{figure}[b]
\centering
\includegraphics[width=0.7\columnwidth]{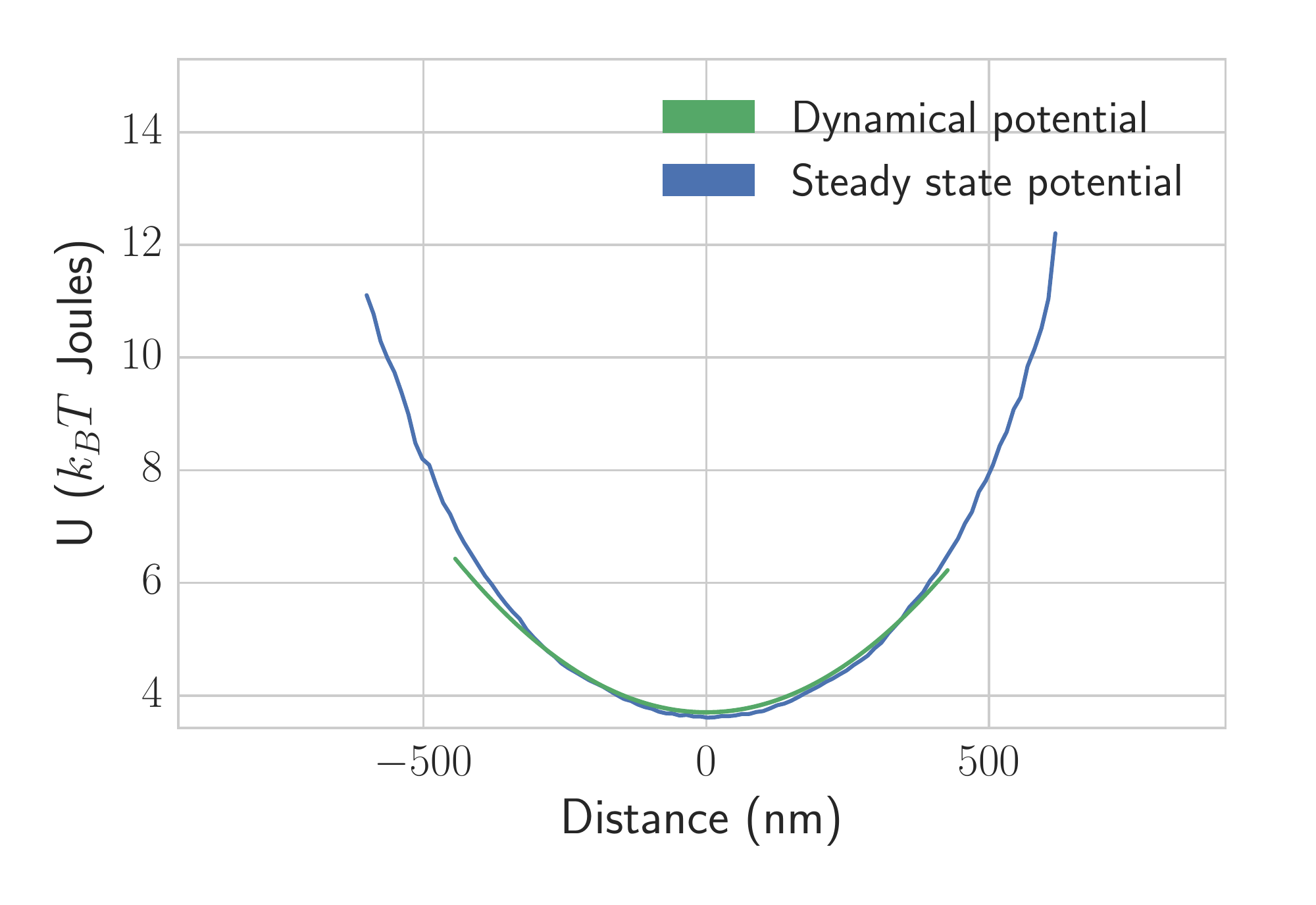}
\caption{{\bf Mass extraction from potentials.} The steady-state, according to equation (\ref{steadystatepotentialequation}), and the kinematic, according to equation (\ref{kinequation}), potentials are fitted to each other with the mass of the particle as the only free parameter. In this way the mass of the particle is extracted from the measured data directly without assumptions otherwise used. The mass extraction from potential has been done for large ($d=11 \mu$m) distances between the trapped particle and the surface.}
\label{fig:mass}
\end{figure}
Since the finite differencing of the particle's time trace, gives acceleration rather than force, we extract $U_{kin}/m$. Thus, on the assumption that the particle is in thermal equilibrium, we equate the two potentials, $U_\text{stst} = U_\text{kin}$, and make mass, $m$, the only free fitting parameter. We then extract the radius $r$ from the mass $m$ on the assumption that the particle is of spherical shape. This assumption is supported by the experimental evidence that the motion in different spatial directions is not coupled. The comparison of the two potentials, $U_\text{stst}$ and $U_\text{kin}$, is shown in figure \ref{fig:mass}, and we extract a particle radius of $r=$ 60nm ($\pm$5nm) for the data shown in this paper.  The error bar of the mass is derived from the fitting error.

%%%%potential extraction

\subsection{Potential extraction}\label{springs}
\textit{Spring functions--} The potentials governing the particle's motion both close ($d=4 \mu$m) to and far away ($d=11 \mu$m) from the surface are extracted by integrating the spring functions at each distance between the nanoparticle trap site and the surface.
\begin{figure}
    \subfloat{\includegraphics[width= 0.35\columnwidth]{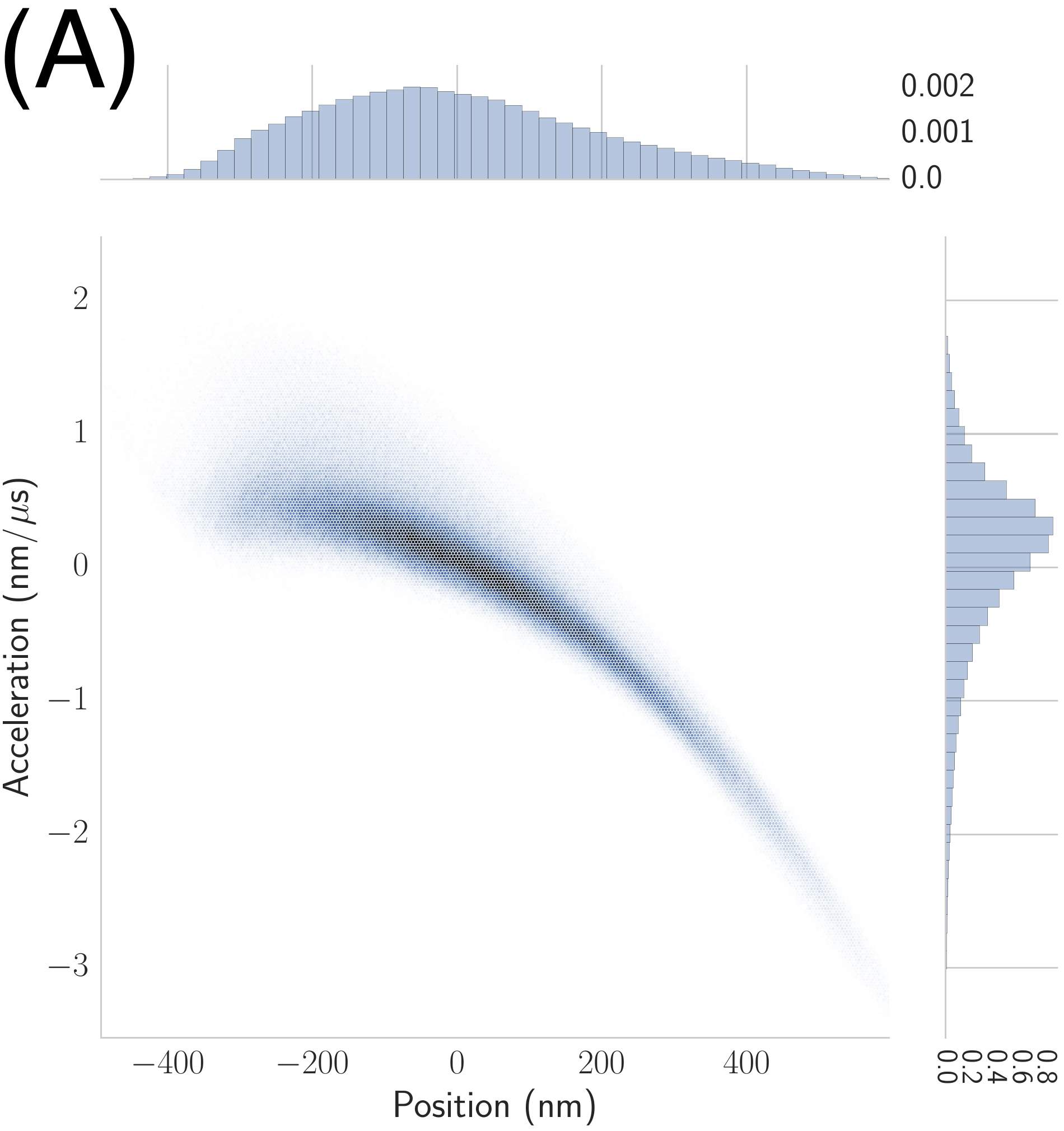}}
    \subfloat{\includegraphics[width= 0.35\columnwidth]{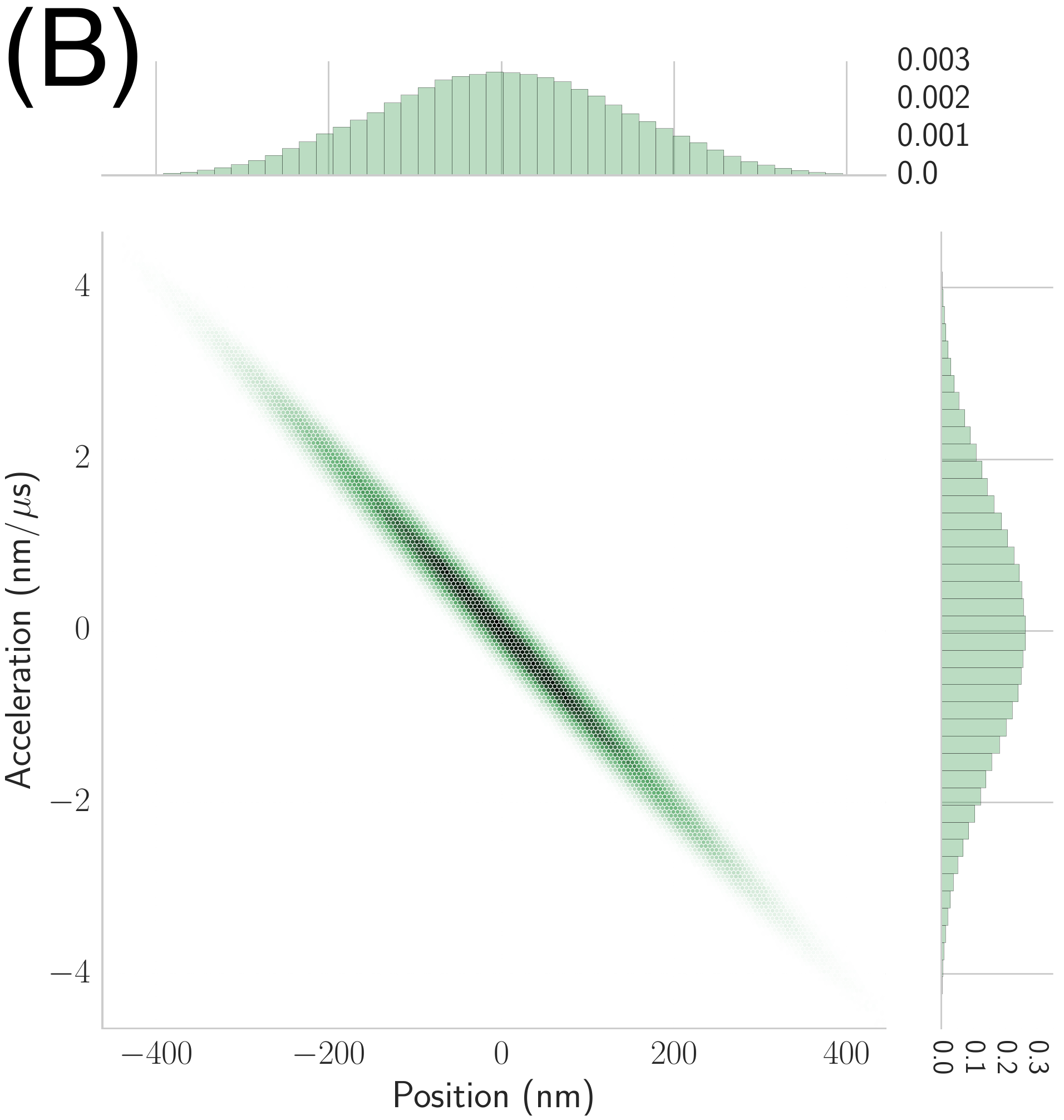}}
    \caption{\textbf{Acceleration - position spring functions.} (A) Spring function for particle close to the surface. (B) Spring function for particle far away from the surface.}
    \label{fig:springs}
\end{figure}
We obtain the spring function, as shown on figure \ref{fig:springs} for the particles motion by numerically differentiating the experimentally obtained time trace of the particles motion with respect to time. The frequency of the particles motion is on the order of 50 kHz, while the sampling frequency of the oscilloscope is 2.5 MHz, giving us about 50 data points per oscillation period and therefore a very good phase space resolution of the particles motion. We can then obtain the potential structure experienced kinematically by the particle by numerically integrating the spring function with respect to space.

{\it  Evaluation of charge $Q$ and distance $d$ from potentials--} As the surface-to-particle distance $d$ becomes smaller the potential experienced by the particle is increasingly perturbed by both optical backscatter and the increasing surface forces from the particle. The changes in the optical power however are generally symmetrical and can be calibrated against the relative frequency drop of the motion in the $x$ and $y$ axis. To extract the non-symmetric surface potential from the total potential we make the ansatz of the Coulomb $(1/x)$-function as a perturbation to the harmonic potential. Then,
\begin{equation}
U_\text{tot} = \frac{1}{2} k x^2 +  U_\text{ic}(d),
\label{Ueffectivefit}
\end{equation}
where $U_\text{tot}$ is the total potential as experienced by the particle and consists of the harmonic optical trap and the surface interaction shown in Eq.~\eqref{mirrorchargepotential}. Here $k$ is the spring constant of the optical trap (for a linear trap), $d$ is the distance between the center of the optical trap and the silicon surface and $Q$ is the charge on the particle. 

$U_{exp}$ is the potential we obtain experimentally, we use the equation:
\begin{equation}
\frac{1}{2} k_\text{new} x^2 = U_\text{exp} -  U_\text{ic}(d),
\end{equation}
where $k_\text{new}$ is a new spring constant, to fit parameters $Q$ and $d$ such that the output of the function becomes a symmetrical potential, which is usually a $kx^2$ function similar to the original, unperturbed, optical function far away from the surface, but with a weaker spring function $k$. Our claim here is that while proximity to the surface does weaken the optical trap it weakens it in a predictable symmetrical way, calibratable by the $x$ and $y$ motional peaks and is not sufficient to explain the anharmonicity experienced by the particle near the surface, whereas a Coulomb image charge model fits well to explain that anharmonicity. 

\subsection{Laser power dependence of trapping potential} Studying the change in spring functions for a trapped particle with different trapping laser powers demonstrates the change is always symmetric, as shown in figure \ref{fig:springs_and_potentials}. We conclude that the observed anharmonicity cannot be explained by a laser power dependent optical effect.
\begin{figure}
\centering
\includegraphics[width=0.7\columnwidth]{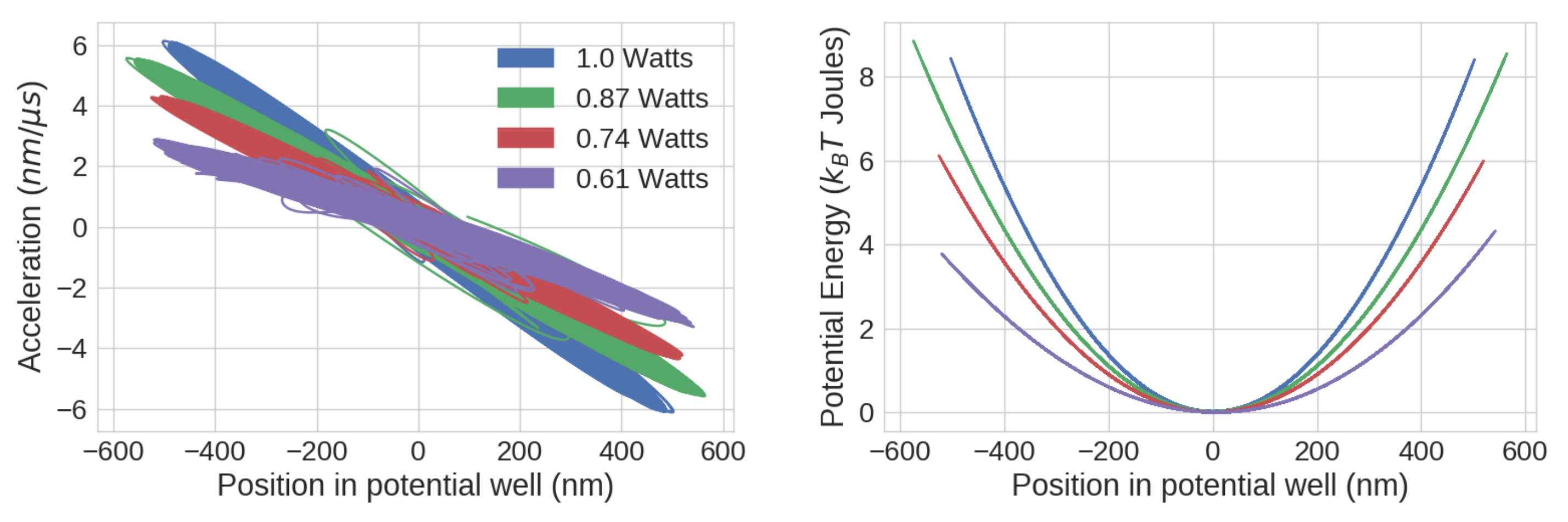}
\caption{{\bf Trapping laser power dependency of spring function and trapping potential.} Despite the change in laser power, the optical potential remains composed of $x^2$ and higher order symmetrical terms. The spring function remains linear. Data have been taken for large distance between particle and surface.}
\label{fig:springs_and_potentials}
\end{figure}

\subsection{Morse potential behaviour} \label{Morsesection}
\textit{Morse potential--} Kinematically the acceleration broadening of the particle's motional states close to the surface is non-intuitive to rectify, in the energetic picture however, the reasoning is somewhat more obvious. For the total potential, the sum of the optical and surface potentials, for the unperturbed and most perturbed examples in the experimental dataset, the limitation on the closest approach to the surface is given when the surface dispersion potential overpowers the optical potential well. Structurally, the overall potential of the system appears similar to a Morse potential, taking the histogram of the particle's energy, the population also follows the expected scaling relation, figure \ref{fig:morse_potential_level_behaviour}.

%\marker{\hrulefill \, essentially all new from here \hrulefill}

\begin{figure}
\includegraphics[width=\columnwidth]{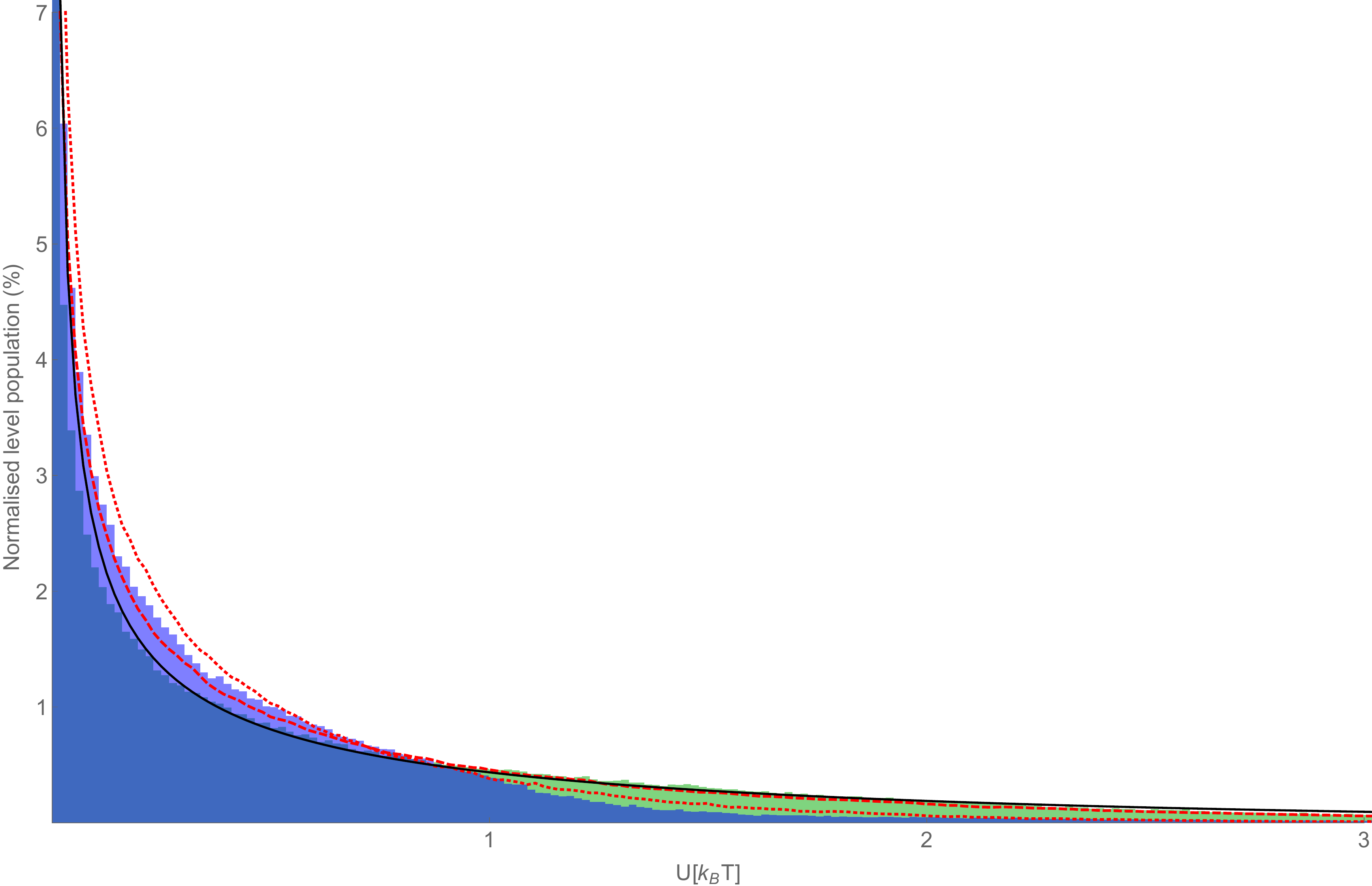}
\caption{Normalised histogram of the particle's potential energy for the cases of an harmonic potential (particle far away from surface, green) and an anharmonic Morse-like potential (particle close to the surface, blue). Shown in red are the results of numerical simulation of the particle's motion for harmonic (dashed) and anharmonic (dotted) potentials with $10 \, 000$ different realisations with randomised amplitudes. The solid black line is the normalised result of equation~\eqref{dtAvg}. The Morse-like potential shows a similar behaviour in energy level scaling as compared to the harmonic case until diverging strongly at a high value of n, which is the expected Morse-like behaviour.}
\label{fig:morse_potential_level_behaviour}
\end{figure}

The shape of the histogram in figure \ref{fig:morse_potential_level_behaviour}. can be understood in the following way. For the unperturbed case, the particle undergoes harmonic motion;
\begin{equation}\label{BasicSHM}
x(t) = x_0 \cos \omega t
\end{equation}
which can then be used in the potential energy $U = \frac{1}{2} k x(t)^2$. Making a histogram of the potential found in this way at evenly-spaced times results in a two-peak shape, in contrast to that shown in figure~\ref{fig:morse_potential_level_behaviour}. This can be described by solving $U = \frac{1}{2} k x(t)^2$ with $x(t)$ given by equation~\eqref{BasicSHM}, giving;
\begin{equation}
t =\frac{1}{\omega} \arccos\left[ \frac{1}{x_0} \sqrt{\frac{2 U}{k}}\right]
\end{equation}
The time spent within any time interval $dt$ will then be given by
\begin{equation}\label{dtEq}
dt = \frac{dt}{dU} dU = - \frac{\text{sgn}(x_0)}{\omega \sqrt{2U} \sqrt{k x_0^2-2U}}dU
\end{equation}
which, subject to proper normalisation, corresponds to the likelihood of the particle being observed within a potential energy between $U$ and $U+dU$. It is clear that $dt$ has maxima at $U=0$ and $U=\frac{1}{2} kx_0^2$, resulting in the two-peak structure discussed above. However, this is not seen in the histogram of energies extracted from the experimental data (fig.~\ref{fig:morse_potential_level_behaviour}), rather a single peak at $U=0$ is found in both the harmonic and anharmonic cases.

This difference comes from the fact that the motion of the particle in the trap is not in fact well-described by equation~\eqref{BasicSHM} --- the experiment takes places at vacuum of $10^{-2}$mbar, meaning the particle's motion is still affected by collisions with the background gas. We incorporate this into the description of the system by assuming that the motion may be taken to be averaged over many trajectories, each with a different amplitude $x_0$, Gaussian-distributed about some mean value $\overline{x}_0$ with a standard deviation $\sigma$. Thus we calculate;
\begin{equation}
\overline{dt} = -\frac{1}{\sqrt{2\pi\sigma^2}}\int^\infty_{\sqrt{2U/k}} dx_0  e^{(x_0-\overline{x}_0)^2/(2\sigma^2)}  \frac{\text{sgn}(x_0)}{\omega \sqrt{2U} \sqrt{k x_0^2-2U}}dU \label{dtAvg}
\end{equation}
with $x_0$ being the integration variable, and $\sigma$ being the only fitting parameter. The result for $\sigma = 100$ nm is shown in figure~\ref{fig:morse_potential_level_behaviour}, alongside the experimental data and a numerical simulation for both harmonic and anharmonic potentials. For the latter case the equations of motion cannot be solved analytically, so there is no equivalent of equation~\eqref{dtAvg} for the anharmonic potential. It is seen that our modelling of the background gas collisions by smearing out the amplitude of the oscillations is consistent with experimental results, especially for relatively high energy. All the results display the required behaviour of having a single peak at $U=0$, which can be intuitively understood from our averaging procedure --- since all the trajectories pass many times through $U=0$ but have different maximum values, the peak of equation~\eqref{dtEq} at $U=\frac{1}{2}kx_0^2$ is suppressed.

%%
%\begin{figure}
%\includegraphics[width=0.5\columnwidth]{appendix_figures/HistvsAnalytic}
%\caption{}
%\label{fig:HistvsAnalytic}
%\end{figure}
%
%
%

\begin{figure}[h!]
\includegraphics[width =0.8\columnwidth]{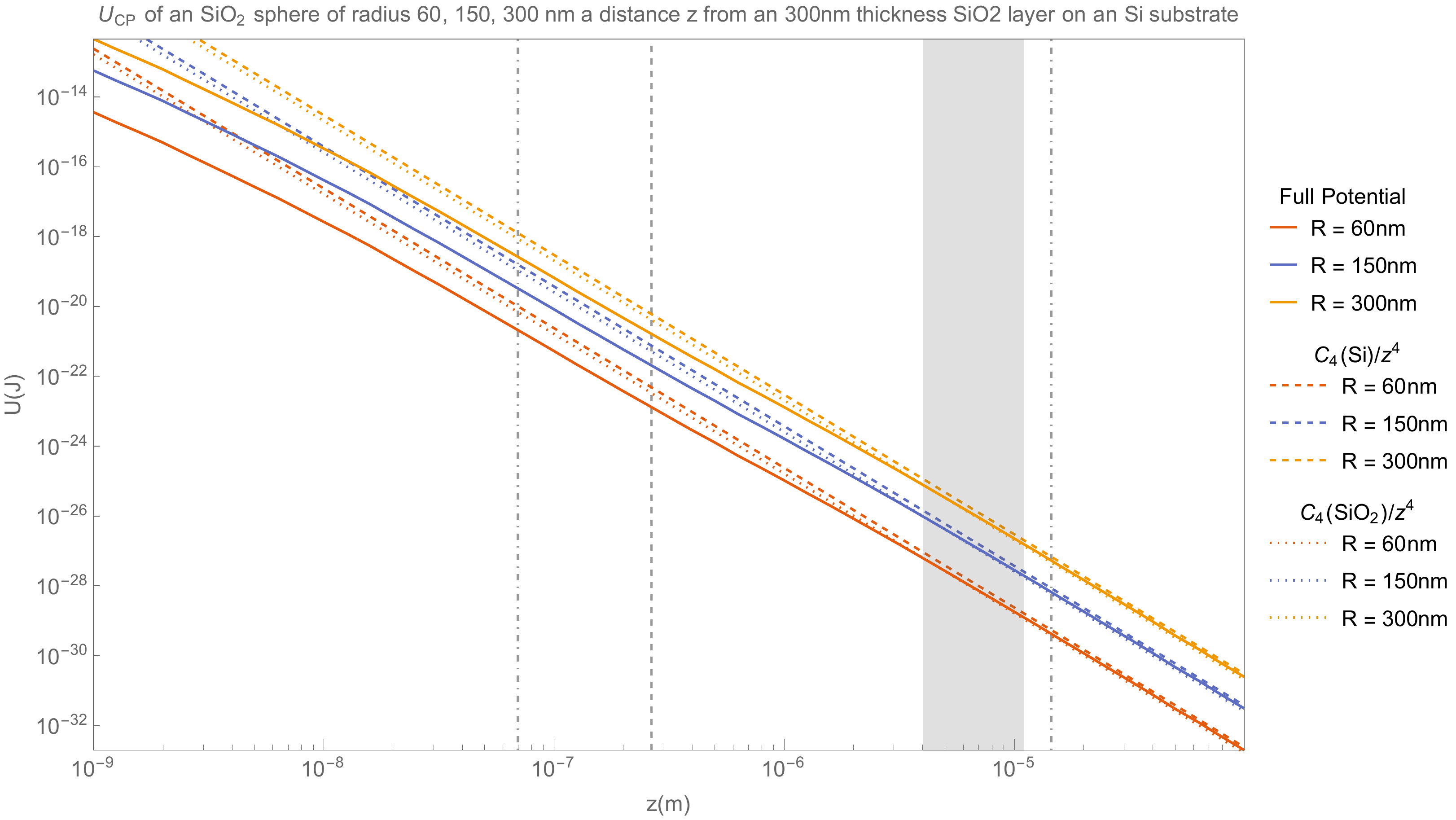}
\caption{Casimir-Polder potential of an SiO$_2$ sphere of various radii near an Si plate. The solid curves are exact results found from numerical integration of \eqref{UGen}, while the dotted curves are from the approximate form equation~\eqref{C4DefSupp}. We also show with dashed lines the result of evaluating equation~\eqref{C4DefSupp} using instead the permittivity of Si for the half-space. The absorption wavelengths of the two media involved are shown as vertical lines (dashed for silicon, dot-dashed for silicon dioxide), while the experimental region of interest ($\sim 4$ to $11\mu$m) is highlighted in grey. It is seen that this experiment is taking place at distances larger than the dominant transition wavelengths of either medium involved (the longer-wavelength silicon dioxide transition is significantly weaker than the shorter-wavelength one, corresponding to a much smaller value of $\omega_p$ in Table \ref{ParamsTable}). Both half-space approximations agree well with the full numerical integration of the layered potential, but in figure~\ref{ErrorPlot} it is seen that closer agreement is found with the silicon dioxide version.} 
\label{PotentialPlot}
\end{figure} 

\subsection{Casimir-Polder potential}

%\marker{\hrulefill \, re-done for layer \hrulefill}

The Casimir-Polder potential $U(z)$ of a particle with polarisibility $\alpha(\omega)$ a distance $z$ from a layer of thickness $L$ and relative permittivity $\epsilon_1(\omega)$, supported by an infinitely deep substrate with relative permittivity $\epsilon_2(\omega)$ is given by \cite{Dzyaloshinskii1961}
\begin{equation}\label{UGen}
 U_{\text{CP}}(z)= \frac{\hbar\mu_0}{8\pi^2}\int_0^\infty \xi^2 \alpha(i\xi) \int_{\xi/c}^\infty d\kappa_0\,  e^{-2\kappa_0 z}\left[R_\text{TE}(\kappa_0,\kappa_1,
\kappa_2)+\left(1-2\frac{\kappa_0^2 c^2}{\xi^2}\right)R_\text{TM}(\kappa_0,\kappa_1\kappa_2)\right]
\end{equation}
where, for either polarization $\sigma$ ($=$TE,TM);
\begin{equation}
R_\sigma = \frac{R^\sigma_{01} + e^{-2 \kappa_1 L}R^\sigma_{12}}{1+e^{-2 \kappa_1L}R^\sigma_{01}R^\sigma_{12}}
\end{equation}
with
\begin{equation}
R^\text{TE}_{ij} = \frac{\kappa_i-\kappa_j}{\kappa_i+\kappa_j} \qquad R^\text{TM}_{ij} = \frac{\varepsilon_j(i\xi)\kappa_i-\varepsilon_i(i\xi)\kappa_j}{\varepsilon_j(i\xi)\kappa_i+\varepsilon_i(i\xi)\kappa_j}
\end{equation}
and 
\begin{align}
\varepsilon_0(i\xi)&=1 &\varepsilon_1(i\xi)&=\epsilon_\text{SiO$_2$} (i\xi)&\varepsilon_2(i\xi)&=\epsilon_\text{Si}(i\xi) & \kappa_i &= \sqrt{\left[\varepsilon_i(i\xi)-1\right]\xi^2/c^2+\kappa_0^2}
\end{align}
A small sphere of radius $R$ may be modelled via the Clausius-Mossotti polarizability
\begin{equation}
\alpha(\omega) =   4\pi  \epsilon_0 R^3  \frac{\varepsilon(\omega)-1}{\varepsilon(\omega)+2}
\end{equation}
where $\varepsilon(\omega)$ is the dielectric function of the material from which the sphere is made, and $\epsilon_0$ is the permittivity of free space. Using this relation, equation~\eqref{UGen} is now a formula whose inputs are the dielectric functions for the media that make up the substrate, layer and sphere, which are all known from experiment. As discussed in the main text, we will eventually approximate this by its large-distance limit near a simple half-space of permittivity $\varepsilon_1(\omega)$, in which case the potential takes on the following form;

\begin{figure}%[h!]
\includegraphics[width =0.8\columnwidth]{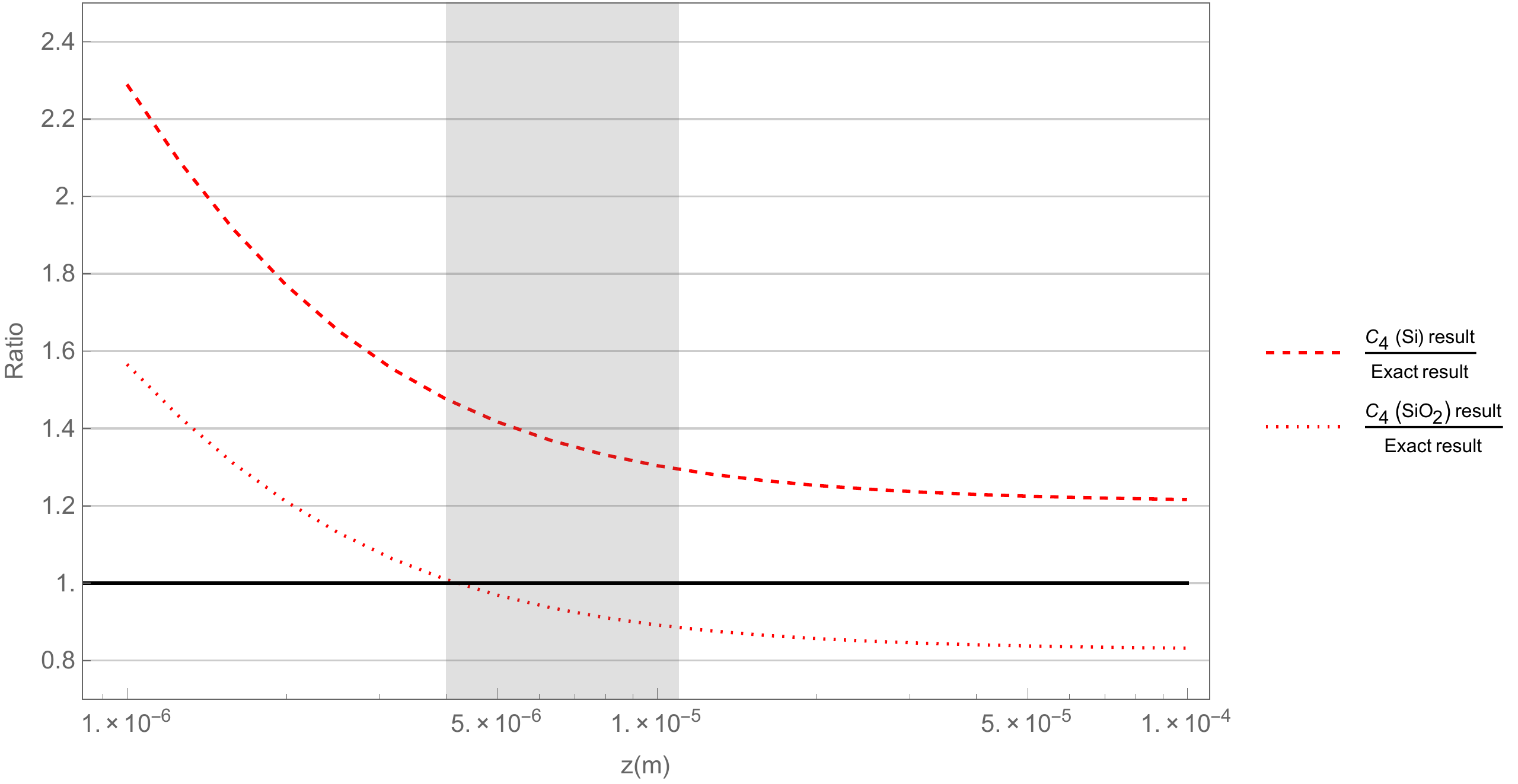}
\caption{Ratio of the approximations shown in Fig.~\ref{PotentialPlot} to the exact numerical result (which is independent of the radius of the sphere). The silicon dioxide result is seen to agree well within the experimental region of interest.  }\label{ErrorPlot}
\end{figure} 

\begin{equation}
U_\text{CP}(z) = -\frac{C_4}{z^4}
\end{equation}   
where $C_4$ is a distance-independent constant defined as \cite{Spruch1993};
\begin{equation}\label{C4DefSupp}
C_4\! = \!\frac{3\hbar c  \alpha(0)}{64\pi^2 \epsilon_0} \!\!\int_1^\infty\! \!\!dv \left(\frac{2}{v^2}\!-\!\frac{1}{v^4}\right) \frac{\varepsilon_1(0) v-\sqrt{\varepsilon_1(0) -1+v^2}}{\varepsilon_1(0) v+\sqrt{\varepsilon_1(0) -1+v^2}}
\end{equation}

In the experiment presented in the main text the surface is silicon and the sphere is silicon dioxide. We model both of these via an $N$-resonance Drude-Lorentz permittivities, defined by;
\begin{equation}
\epsilon(\omega) = 1+ \sum_{i=1}^N \frac{\omega_{p,i}^2}{\omega_{T,i}^2-\omega^2+i \gamma_i \omega}
\end{equation}
Here $\omega_{p,i}$ is the plasma frequency, $\omega_{T,i}$ is the transition frequency and $\gamma_i$ is the damping frequency, each for the $i$th resonance of the dielectric function. For silicon we use a single-resonance model, and for silicon dioxide we use a two-resonance model, with parameters from \cite{Palik1985} shown in Table~(\ref{ParamsTable}).
\begin{table}[h!]
\centering
        \begin{tabular}{| l | c | c | c | c |}
            \hline
                      &  $i$ & $\omega_{p,i}$ & $\omega_{T,i}$ & $\gamma_i$ \\                \hline
               Si &  1 & 23  &  7.1 & 0.98\\                \hline
            \multirow{2}{*}{SiO$_2$} & \multicolumn{1}{c|}{1} & \multicolumn{1}{c|}{0.17} & \multicolumn{1}{c|}{0.13} & \multicolumn{1}{c|}{0.043} \\
                            & \multicolumn{1}{c|}{2} & \multicolumn{1}{c|}{29} & \multicolumn{1}{c|}{27} & \multicolumn{1}{c|}{8.1} \\\hline
        \end{tabular}
    \caption{Drude-Lorentz parameters for silicon and silicon dioxide (all values in units of $10^{15}$rad/s)}
    \label{ParamsTable}
\end{table}
Using these parameters in Eq.~\eqref{C4DefSupp} we find the dispersion constant $C_4$ for our particular setup
\begin{equation}\label{CResults}
C_4= (7.60\times10^{-28}\text{Jm})\cdot R^3 \quad 
\end{equation}
As a consistency check we evaluate Eq.~\eqref{UGen} numerically over both distance regimes, the results of this alongside the asymptotic long-distance result according to equation~\eqref{CResults} are shown in figure~\ref{PotentialPlot} for a range of sphere sizes. 

\end{document}